\documentclass[sigconf]{acmart}

\setcopyright{none} 
\renewcommand\footnotetextcopyrightpermission[1]{}

\usepackage{graphicx}
\usepackage{array}
\usepackage{tabularx}
\usepackage{siunitx}
\usepackage{xcolor}
\usepackage{multirow}
\usepackage{makecell}
\usepackage{comment}
\usepackage{framed}

\definecolor{shadecolor}{gray}{0.95}

\newenvironment{rqbox}{%
  \par\noindent
  \setlength{\FrameRule}{0.4pt}
  \setlength{\FrameSep}{6pt}
  \vspace{0pt}%
  \begin{shaded*}\noindent
}{%
  \end{shaded*}%
  \vspace{0pt}%
}

\begin{document}

\title{Role of CI Adoption in Mobile App Success: An Empirical Study of Open-Source Android Projects}

\author{Xiaoxin Zhou}
\affiliation{
  \department{Faculty of Information}  
  \institution{University of Toronto}
  \city{Toronto}
  \state{Ontario}
  \country{Canada}
}
\email{xiaoxin.zhou@utoronto.ca}

\author{Taher A. Ghaleb}
\orcid{0000-0001-9336-7298}
\affiliation{
  \department{Department of Computer Science} 
  \institution{Trent University}
  \city{Peterborough}
  \state{Ontario}
  \country{Canada}
}
\email{taherghaleb@trentu.ca}

\author{Safwat Hassan}
\orcid{0000-0001-7090-0475}
\affiliation{
  \department{Faculty of Information}
  \institution{University of Toronto}
  \city{Toronto}
  \state{Ontario}
  \country{Ontario}
}
\email{safwat.hassan@utoronto.ca}

\begin{abstract}

Mobile apps face strong pressure for fast and reliable updates. Continuous Integration (CI) helps automate builds, tests, and releases, but its impact on mobile development remains underexplored. 
Despite the widespread use of CI, little is known about how it affects development activity, release speed, and user-facing outcomes in mobile projects. 
Existing studies mostly focus on CI adoption in general-purpose software, providing limited insight into mobile-specific dynamics, such as app store visibility and user engagement. 
In this paper, we analyze open-source Android apps to (1) compare CI adopters and non-adopters, (2) characterize adoption patterns using activity and bug metrics, and (3) assess pre/post adoption changes and user-facing outcomes. 
We observe that CI adopters are larger and more active, with faster and more regular releases. CI adoption is concentrated in integration- and reliability-intensive categories (e.g., finance and productivity) and is associated with higher Google Play Store engagement (more downloads and reviews) without lower ratings. 
Overall, CI adoption aligns with practices that support sustained delivery, higher project visibility, and stronger user engagement in mobile ecosystems.
\end{abstract}



\makeatletter
\renewcommand{\shortauthors}{}%
\renewcommand{\shorttitle}{}%
\let\@acmConference\@empty
\let\@acmYear\@empty
\let\@acmISBN\@empty
\let\@acmDOI\@empty
\let\@acmPrice\@empty

\fancypagestyle{emptyheadstyle}{%
  \fancyhf{}
  \fancyhead[L]{}
  \fancyhead[C]{}
  \fancyhead[R]{}
  \fancyfoot[C]{\thepage}
  \renewcommand{\headrulewidth}{0pt}%
  \renewcommand{\footrulewidth}{0pt}%
}
\pagestyle{emptyheadstyle}
\thispagestyle{emptyheadstyle}

\def\@copyrightspace{}%
\makeatother

\maketitle

\section{Introduction}
\label{sec:introduction}
Mobile applications (hereafter referred to as mobile \textit{apps}) are an integral part of daily life. To succeed, developers must rapidly deploy new releases that deliver requested features or fix reported bugs~\cite{DBLP:journals/ese/McIlroyAH16}. Continuous Integration (CI) has become a widespread practice in software development, including mobile apps, helping to automate building, testing, packaging, and deployment~\cite{fowler2006continuous,Difference_Between_CI_CD,Difference_Between_CI_CD2}.
Previous work studied practices to improve build performance, such as reducing build execution time and minimizing build failures~\cite{ghaleb2019empirical,ghaleb2019studying,bouzenia2024resource,jin2021helped}. Other research has examined CI adoption in Android apps, focusing primarily on configuration and maintenance practices. 
However, little is known about how CI adoption in mobile apps affects development productivity, release speed, and user-facing success metrics~\cite{DBLP:conf/esem/WangZXY23, DBLP:conf/kbse/LiuSZL0022,ghaleb2025android}.

In this paper, we examine the association between CI adoption and various aspects of software development in 2,542 open-source Android projects. Specifically, we investigate how CI usage is associated with development practices, release patterns, and app success. To this end, we address the following research questions (RQs):

\vspace{3pt}
\noindent \textbf{RQ1: How do CI adopters differ from non-adopters?}
We compare Android projects with and without CI adoption. CI adopters are larger and more active, with higher source lines of code (+24\%), more contributors (+260\%), more commits (+280\%), and faster releasing (+41\% tags/week, 18\% shorter inter-tags). 
Both have similar Google Play Store listing rates (\(\approx33\%\)), but CI adopters have \(\sim5\times\) more downloads and \(\approx2.2\times\) more reviews, especially in utility and integration-heavy categories.

\vspace{3pt}
\noindent \textbf{RQ2: What are the distinct patterns of CI adopters?}
We focus on CI adopters with labeled bug issues (\(n=442\)) and normalize activity metrics. Among 442 CI adopters with labeled bugs, we find that faster releasing does not increase bug density. 
The \textit{“Best”} group has short tag lifetimes (8 days) and low bug density (0.23 issues/KLoC), while the \textit{“Worst”} group has longer cycles (32 days) and higher bug density (2.73 issues/KLoC). 

\vspace{3pt}
\noindent \textbf{RQ3: What are the evolution characteristics of CI adoption?}
A before/after analysis of 172 projects shows post-adoption increases in commits (+59\%), PRs opened (+75\%), and PRs merged (+72\%), but merges slowed (median ratio = 1.6). 
The User-facing metrics on Google Play show no significant improvement.

\vspace{3pt}
\noindent \textbf{RQ4: What explains the variation in before- and after-adoption performance?}
Clustering 61 projects by before/after change patterns reveals three CI adoption trajectories: (1) higher activity with shorter releases, (2) lower activity with longer releases, and (3) higher activity with longer releases, reflecting diverse outcomes shaped by project practices and pipeline setups.

\vspace{3pt}
\noindent
Overall, this paper makes the following contributions.
\begin{itemize}
  \item A publicly available dataset~\cite{our_replication_package} linking CI adoption, repository metrics, and Google Play Store outcomes.
  \item Models (Random Forest and $k$-means clustering) that reveal patterns in CI adoption across different metrics (e.g., commits, pull requests (PRs), release speed, bug density).
  \item A dataset and analysis of how CI adoption affects development throughput, release speed, and user outcomes.
  \item Insights into the variability of post-adoption performance, showing that higher activity does not necessarily correlate with faster releases.
\end{itemize}

We organize the rest of this paper as follows. 
Section~\ref{sec:Methodology} presents our study setup. Section~\ref{sec:Results} discusses our empirical analyses and results.
Section~\ref{sec:Implications} discusses the implications of our work. 
Section~\ref{sec:Threats_to_Validity} describes possible threats to the validity of our study.
Section~\ref{sec:Related_Work} reviews work related to CI and mobile apps.  
Finally, Section~\ref{sec:Conclusion} concludes our paper and suggests future directions.

\section{Study Setup}
\label{sec:Methodology}
\begin{figure*}[!htbp]
  \centering
  \includegraphics[width=0.995\linewidth]{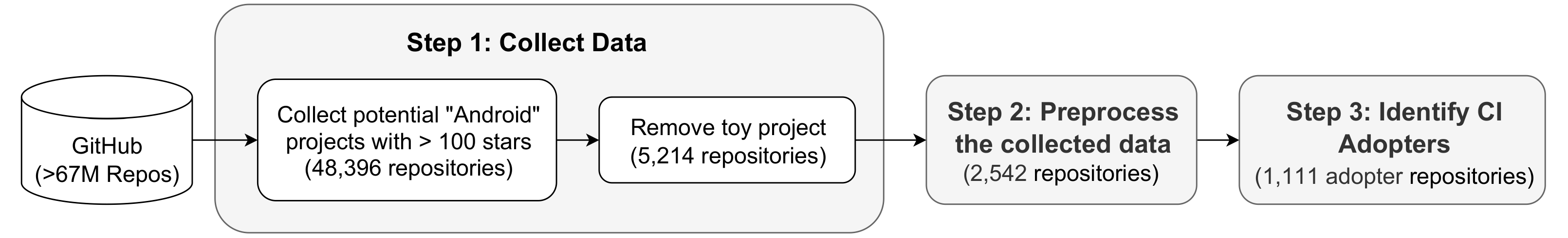}
  \vspace{-10pt}
  \caption{Data collection and preprocessing pipeline for CI analysis.}
  \Description{Overview of the study methodology for data collection and preprocessing pipeline for CI analysis.}
  \label{fig:data-collection-overview}
\end{figure*}

This section presents the setup of our study dataset. We explain how we collect and preprocess the data to address our research questions. We break the analysis and data preprocessing into three steps: 1) Collect Data, 2) Data Preprocessing, and 3) CI Data Analysis, as shown in Figure~\ref{fig:data-collection-overview} and described below. We make our final clean dataset publicly available on Figshare~\cite{our_replication_package}.

\subsection{Data Collection}
For this study, we collect a dataset of Android repositories from GitHub (in July 2025) using a source that has been widely adopted in prior research~\cite{ghaleb2025android},

which focuses on open-source Android apps and provides publicly accessible information on CI configurations. Building on this foundation, we systematically gather and curate the repositories following the steps outlined below.

\subsubsection{\textbf{Collecting potential ``Android'' repositories}}
We collect Android projects from GitHub using the official GitHub REST API\footnote{\url{https://docs.github.com/en/rest}}. GitHub hosts more than 67 million repositories in total. To narrow down our dataset, we search for repositories with the \texttt{android} topic\footnote{\url{https://github.com/topics/android}}, or the word \texttt{android} is mentioned in the project title, description, or read-me files. 
We also search for projects with ``Java'', ``Kotlin'', or ``Dart'' as their primary programming language.
In addition, we work on repositories with more than 100 stars to ensure that our dataset consists of mature and actively used projects. We follow established guidelines used in prior studies~\cite{9978190,10.1007/s10664-024-10497-x}, which also use a 100-star threshold to ensure selecting popular projects and avoid toy ones.
This step, conducted in July 2025, resulted in the identification of 48,396 repositories.

\subsubsection{Removing toy projects: }
To further refine the dataset, we follow prior work~\cite{ghaleb2025android} to exclude projects without a user interface by verifying the presence of an \texttt{AndroidManifest.xml} file containing at least one \texttt{activity} tag. 
In Android, activities define the screens of an application; therefore, the existence of a manifest file with an activity indicates that the project represents a functional app rather than a demo or library project. 

We then exclud toy projects (e.g., tutorials, sample apps, and library projects) that mimic the structure of real Android apps but do not represent actual mobile applications~\cite{ghaleb2025android}.
This filtration step ensures that the selected projects are indeed Android apps, thereby enhancing the accuracy and relevance of our dataset for analysis. 
At the end of this step, we obtain 5,214 repositories.

\subsection{Data Preprocessing}
To ensure that our analysis captures meaningful signals of continuous integration and continuous delivery (CI) activity, we first preprocess the dataset of GitHub repositories. Our goal is to focus on repositories that demonstrate evidence of versioning, issue tracking, and sustained development over time. The following filtering steps are applied. We should note that we update our dataset in October 2025 to incorporate the most recent repository information.

\subsubsection{Remove repositories with no tags: } 
In GitHub repositories, tags can serve as mechanisms for tracking the history of software updates. 

Since our analysis focuses on the speed of updates, we examine the publication dates associated with the project tags to estimate update frequency and, indirectly, contributor efficiency. 
To avoid noise from inactive or abandoned projects, we apply a filter requiring that each repository contain at least one tag (i.e., more than zero updates). After implementing this filtering process, we identify 2,690 repositories suitable for analysis.

\subsubsection{Remove repositories with no issues: }
In GitHub repositories, contributors and other interested users can report problems, bugs, or feature requests by creating issues. These issues serve as a communication channel between users and project maintainers, providing valuable feedback about software quality and functionality. 
To reduce ambiguity and focus only on repositories with observable issue-tracking activity, we apply a filter to exclude all repositories with zero reported issues. Combined with our earlier filter that excludes repositories without tags, this process left us with 2,560 repositories. This ensures that the dataset reflects projects with both measurable update histories and at least minimal evidence of issue reporting, allowing for a more meaningful analysis of development activity and code quality.

\subsubsection{Remove repositories with no activities:}
For this particular analysis, we use October 1, 2025, as the cutoff date and select repositories that show activity, defined as having at least one commit, within the 90 days prior. This approach focuses the study on recently active repositories, ensuring that our conclusions are timely and relevant. After applying this criterion, we identify 2,542 Android repositories for inclusion in our study. The 90-day activity window follows prior work, including Moriconi et al.~\cite{11025655} and Khelifi et al.~\cite{10992408}, which use a 90-day period to ensure selecting actively maintained projects.

\subsection{Identifying CI adopters}
In this phase, we analyze the final dataset of \textbf{2,542} Android repositories to (i) identify which projects adopt CI and which do not, and (ii) determine which projects are published on the Google Play Store. 
Following prior work~\cite{ghaleb2025android}, we identify projects that adopt CI by searching for configuration files associated with popular CI services (e.g., GitHub Actions). Each service defines its pipeline through one or more YAML files placed at specific paths in the repository. Table~\ref{tab:cicd-yml} summarizes these services and their corresponding file patterns. Using this mapping, we classify repositories into two groups: \emph{CI adopters}, which contain at least one recognized configuration file, and \emph{non-adopters}, which do not. As a result of this classification, we identify 1,111 CI adopters repositories and 1,431 non-adopters repositories that did not.

\begin{table}[htbp]
  \centering
  \caption{List of CI services with their corresponding YAML file path patterns~\cite{ghaleb2025android}.}
  \label{tab:cicd-yml}
  \vspace{-7pt}
  \begin{tabular}{@{}p{2cm}l@{}}
    \toprule
    \textbf{CI service} & \textbf{Path (regular expression)} \\
    \midrule
    GitHub Actions  & {\ttfamily \detokenize{r'\.github/workflows/.*\.(yml|yaml)}\$'} \\
    Travis CI       & {\ttfamily \detokenize{r'\.travis\.yml}\$'} \\
    CircleCI        & {\ttfamily \detokenize{r'(\.circleci/config\.yml|circle\.yml)}\$'} \\
    GitLab CI    & {\ttfamily \detokenize{r'\.gitlab-ci\.yml}\$'} \\
    Azure Pipelines & {\ttfamily \detokenize{r'azure-pipelines\.yml}\$'} \\
    AppVeyor        & {\ttfamily \detokenize{r'(\.appveyor\.yml|appveyor\.yml)}\$'} \\
    Bitbucket       & \texttt{\detokenize{r'bitbucket-pipelines\.yml}\$'} \\
    \bottomrule
  \end{tabular}
  \vspace{-8pt}
\end{table}

In addition to contrasting CI adopters with non-adopters, we incorporate an app-distribution signal by verifying whether each repository corresponds to an application listed on the Google Play Store. Specifically, for every repository page, we apply a regular expression to extract candidate Google Play Store URLs from project metadata (e.g., README). We then programmatically request each candidate URL and mark the app as ‘exists’ only if the response indicates a valid listing (e.g., HTTP 200 and presence of canonical Play Store markers), thereby minimizing false positives from soft 404s or region-restricted entries.

\section{Empirical Analysis and Results}
\label{sec:Results}
In this section, we present the results of our empirical analyses across four research questions (RQs). For each RQ, we describe the motivation, outline our approach, and summarize the key findings.

\subsection{RQ1: How do CI adopters differ from non-adopters?}
\subsubsection{\textbf{Motivation}}
CI has been widely adopted in software development as a practice that can support more efficient workflows and maintainable code. However, it is not clear how repositories that adopt CI differ from those that do not in real-world open-source Android projects. Understanding these differences is important for characterizing CI adoption patterns and identifying traits associated with active or well-maintained projects. This RQ addresses these differences by examining the observable characteristics of CI adopters compared to non-adopters.

\subsubsection{ \textbf{Approach:}}
We collect and analyze repository-level metrics, including the number of issues, contributors, source lines of code (SLOC; computed using the open-source tool \texttt{sloc}\footnote{\url{https://github.com/flosse/sloc}}), project lifetime, median tag interval, issue density (issues per KLoC), commit frequency, and release/tag frequency, using a 90-day activity window preceding October~1,~2025.

\vspace{3pt}
\noindent\textbf{\textit{Step 1: CI Data Collection.}} We detect and extract CI configurations (e.g., workflow files and run histories) and align them with each repository’s activity timeline.

\vspace{3pt}
\noindent\textbf{\textit{Step 2: Data Preprocessing.}} We deduplicate records, normalize units, and filter out repositories missing critical information (i.e., those without issue histories, with effectively zero lifetime, such as projects created via a single bulk import) or lacking tag/release records. Repositories with incomplete or inconsistent metadata are excluded to ensure valid comparisons.

\vspace{3pt}
\noindent\textbf{\textit{Step 3: Classification and Google Play Store Linking.}} Repositories are classified into CI adopters (CI) and non-adopters (Non-CI), which we sometimes refer to interchangeably as CI vs. non-CI for brevity. We link each GitHub repository to its corresponding Google Play Store app (when available) to study user engagement metrics. For this analysis, we restrict our attention to repositories that are active within the 90 days preceding October~1,~2025, ensuring that engagement measures reflect recent activity. Additionally, for category-level analyses, we assign each app to its primary Google Play Store category (e.g., Finance) and focus only on categories containing at least five repositories to improve statistical reliability.

\vspace{3pt}
\noindent\textbf{\textit{Step 4: Complexity and Evolution Analysis.}} Using the filtered dataset ($df_{\text{clean}}$), we compute four repository-level metrics to compare CI adopters and non-adopters, as follows:

\begin{enumerate}
    \item \textit{Project lifetime.} For each repository ($r$), we take the project creation timestamp ($\text{created\_at}_r$) and the last commit timestamp ($\text{last\_commit}_r$), and compute
\[
\text{ProjectLifetime}_r = \text{last\_commit}_r - \text{created\_at}_r \quad (\text{in days}).
\]
This measures the active development span observed in history.

\item 
\textit{Median tag (release) lifetime.} For each repository, we collect all release tag dates $\{t_1,\dots,t_n\}$, sort them by time, form consecutive gaps $\Delta_i=t_{i}-t_{i-1}$ (days), and define
\[
\text{MedianTagLifetime}_r = \operatorname{median}\{\Delta_2,\ldots,\Delta_n\}.
\]
If a repository has $n<2$ tags, the quantity is undefined and we remove these repositories from our analysis. This captures a typical release speed while being robust to outliers.

\item 
\textit{Issues density.} We proxy issue pressure by normalizing the total number of issues by code size (source lines of code, SLOC):
\[
\text{IssuesDensity}_r=\frac{\text{NoIssues}_r}{\text{SLOC}_r}\times 1000
\quad\text{(issues per KLOC)}.
\]
If $\text{SLOC}_r=0$, we set the metric to $0$ and  remove these repositories from our analysis.

\item 
\textit{Tag (release) frequency.} We measure how often releases occur over the observed lifetime. Let $\text{NoTags}_r$ be the number of tags and $\text{ProjectLifetime}_r$ in days; then
\[
\text{ReleaseFreqWeek}_r=\frac{\text{NoTags}_r}{\text{ProjectLifetime}_r}\times 7
\quad\text{(tags/week)}.
\]
If $\text{ProjectLifetime}_r=0$, we set $0$ and remove these repositories from our analysis.
\end{enumerate}

\vspace{3pt}
\noindent\textbf{\textit{Step 5: Statistical comparison.}} To assess whether the distributions of these metrics differ significantly between CI adopters and non-adopters, we apply the Mann–Whitney U test~\cite{Mann1947}, a nonparametric test for comparing two independent groups.

\vspace{3pt}
\noindent\textbf{\textit{Step 6: Preprocessing and Visualization.}} We remove $\pm\infty$, drop rows with \texttt{NaN} in computed fields, and (when plotting) apply $\log_{10}$ transforms with small constants to avoid $\log(0)$ issues (e.g., $+1$ for counts, $+0.001$ for densities/frequencies). We visualize each metric via split violin plots (CI vs.\ non-CI) with median overlays, using consistent hue ordering and sample sizes in the legend.

\subsubsection{\textbf{Findings:}} We present results from a comparative analysis of CI adopters and non-adopters across multiple repository metrics.

\begin{figure}[!htbp]
  \centering
  \vspace{-5pt}
  \includegraphics[width=\linewidth]{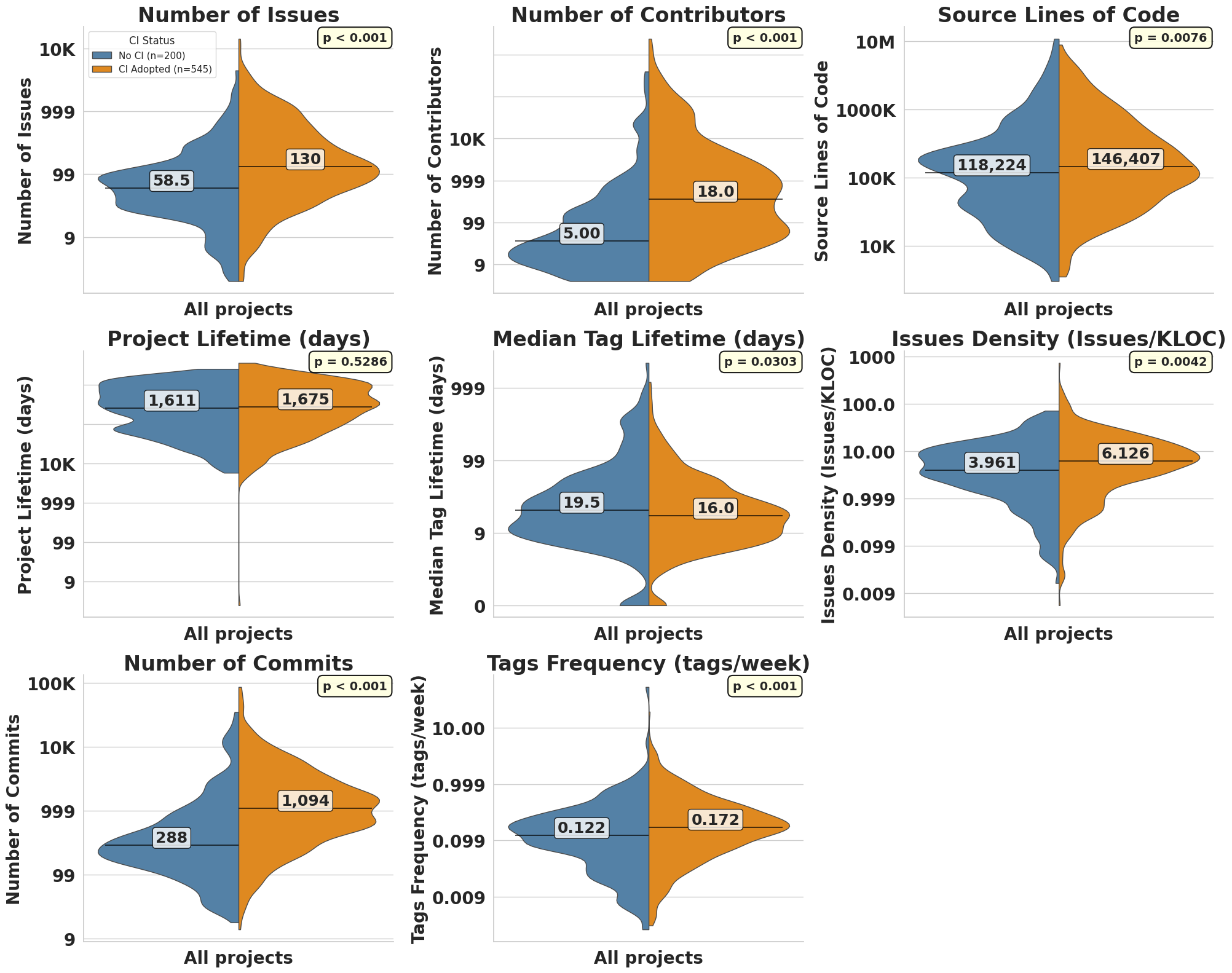}
  \vspace{-10pt}
  \caption{The 90-day activity repository metrics comparison: CI vs Non-CI projects.}
  \Description{A comparison of repository-level metrics (e.g., number of contributors and number of commits) between CI adopters and non-CI adopters.}
  \label{fig:repository_metrics_comparison}
  \vspace{-7pt}
\end{figure}

\vspace{4pt}
\noindent\textbf{Observation 1.1: CI adopters are more mature projects in terms of size, contributors, and project age.}
Figure~\ref{fig:repository_metrics_comparison} shows log-scaled beanplots comparing Non-CI ($n{=}200$) and CI adopters ($n{=}545$), and Table~\ref{tab:ci_comparison_medians} summarizes key repository-level metrics, reporting medians for each group, the absolute difference (Diff.), and relative change ($\Delta$ \%) across metrics including SLOC, number of contributors, project age, commits, release frequency, median tag interval, and issue density. CI adopters are consistently larger and more active: median SLOC is $+23.8\%$ (p = 0.008), number of contributors $+260.0\%$ (p = 0.001), and project age slightly longer $+4.0\%$ (p = 0.529). Development intensity is higher, with $+279.9\%$ more commits (p = 0.001), $+41.6\%$ more tags/week (p = 0.001), and a shorter median inter-tag time $-17.9\%$ (p = 0.030). Issue density is also higher ($+54.6\%$, p = 0.008), likely reflecting greater usage, richer reporting practices, or domain complexity rather than lower quality.

CI adopters stand out as larger, more active, and faster-releasing projects. The higher issue density appears driven by scale and observability rather than poor quality, highlighting that CI adoption correlates with stronger delivery discipline and greater development capacity in mobile software contexts.

\begin{table}[ht]
\caption{Repository Metrics: CI vs.\ Non-CI (Medians; $n_{\text{non-CI}}{=}200$, $n_{\text{CI}}{=}545$).}
\label{tab:ci_comparison_medians}
\vspace{-7pt}
\centering
\resizebox{\columnwidth}{!}{%
\begin{tabular}{@{} l r r r r @{}}
\toprule
\textbf{Metric} & \textbf{Non-CI} & \textbf{CI} & \textbf{Diff.} & \textbf{$\Delta$ (\%)} \\
\midrule
Number of Issues                        & \num{58.5}        & \num{130}       & +\num{71.5}     & +122.2 \\
Number of Contributors                  & \num{5}         & \num{18}        & +\num{13}     & +260.0 \\
Source Lines of Code                    & \num{118224}    & \num{146407}    & +\num{28183}  & +23.8  \\
Project Lifetime (days)                 & \num{1611}      & \num{1675}      & +\num{64}     & +4.0   \\
Median Tag Lifetime (days)              & 19.5            & 16.0            & -3.5          & -17.9  \\
Issues Density (Issues/KLOC)            & 3.961           & 6.1           & +2.2        & +54.7  \\
Number of Commits                       & \num{288}       & \num{1094}      & +\num{806}    & +279.9 \\
Tags Frequency (tags/week)              & 0.12           & 0.17           & +0.05        & +41.0  \\
\bottomrule
\end{tabular}}
\Description{Medians for eight metrics; CI projects are larger, commit and release more frequently; issue density is higher.}
\vspace{-1pt}
\end{table}

\vspace{7pt}
\noindent\textbf{Observation 1.2: CI-adopting apps show higher user engagement on Google Play Store.}

\begin{figure}[!htbp]
  \centering
  \includegraphics[width=\linewidth]{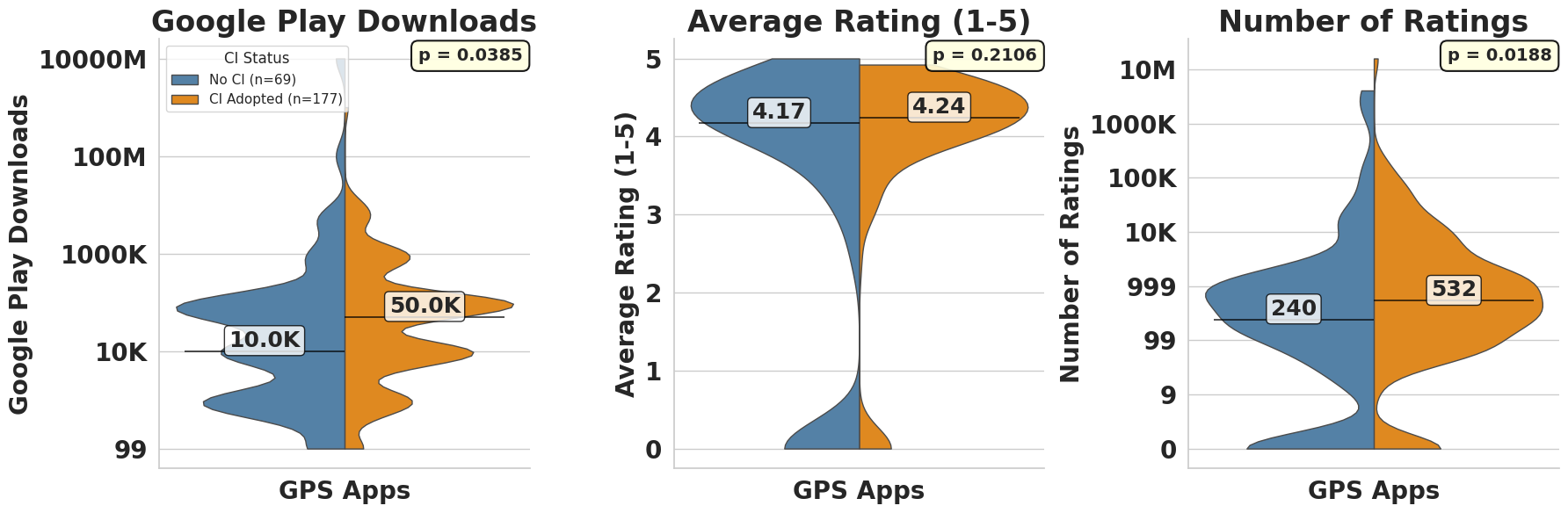}
  \caption{The 90-day activity Google Play Store (GPS) metrics for CI adopters vs Non-CI adopters.}
  \Description{A comparison of Google Play Store (GPS) metrics between CI adopters and non-CI adopters.}
  \label{fig:google_play_store_metrics}
  \vspace{-7pt}
\end{figure}

Among repositories active within the 90 days preceding October~1,~2025, the share with a corresponding Google Play Store listing is similar for CI adopters and non-adopters (CI: 177/545 = 32.5\%, Non-CI: 69/200 = 34.5\%), indicating that CI adoption does not increase the likelihood of being listed. Instead, the advantages appear \emph{post-listing}: CI adopters exhibit substantially higher user engagement. Figure~\ref{fig:google_play_store_metrics} and Table~\ref{tab:gps_comparison_medians} summarize metrics across \textbf{\textit{246}} repositories (Non-CI $n=69$, CI $n=177$). Median downloads for CI adopters are \emph{five times} higher (+400.0\%, p = 0.039), and median ratings are over \emph{double} (+121.7\%, p = 0.019). Average star ratings are slightly higher for CI adopters (+1.7\%), but not statistically significant (p = 0.211). These results suggest that CI-adopting apps achieve stronger uptake and user interaction post-publication.

\begin{table}[ht]
\caption{Google Play Store (GPS) Metrics: CI vs.\ Non-CI (Medians; $n_{\text{non-CI}}{=}69$, $n_{\text{CI}}{=}177$).}
\label{tab:gps_comparison_medians}
\vspace{-7pt}
\centering
\resizebox{\columnwidth}{!}{%
\begin{tabular}{@{} l r r r r @{}}
\toprule
\textbf{Metric} & \textbf{Non-CI} & \textbf{CI} & \textbf{Diff.} & \textbf{$\Delta$ (\%)} \\
\midrule
GPS Downloads              & \num{10000} & \num{50000} & +\num{40000} & +400 \\
GPS Avg Rating (1--5)      & 4.17       & 4.24       & +0.07       & +1.7   \\
GPS Number of Reviews      & \num{240}   & \num{532}   & +\num{292}   & +121.7 \\
\bottomrule
\end{tabular}}
\Description{Median Google Play Store metrics comparing CI adopters vs non-adopters: CI apps have higher downloads and more ratings, with a slightly higher average rating.}
\vspace{-7pt}
\end{table}

\vspace{4pt}
\noindent\textbf{Observation 1.3: CI adoption skews toward utility-style apps, particularly Tools and Productivity, and dominates categories where reliability and frequent updates are critical.}
Table~\ref{tab:gps_category_breakdown_small} summarizes categories with at least five repositories (as described in Step~3). CI adopters are the majority in every category, averaging $\approx$71.5\% overall (153/214).
CI adoption is most prevalent in categories with frequent updates or reliability-critical features: \emph{Finance} (10/10, 100\%), \emph{Maps \& Navigation} (5/5, 100\%), \emph{Social} (6/6, 100\%), \emph{Communication} (10/12, 83.3\%), and \emph{Personalization} (7/8, 87.5\%). Utility-heavy categories like \emph{Tools} (64/96, 66.7\%) and \emph{Productivity} (21/33, 63.6\%) also show clear majorities. Entertainment- and media-oriented categories are closer to parity: \emph{Photography} (2/5, 40\%) and \emph{Health \& Fitness} (3/5, 60\%). Overall, the results can be explained as CI adoption is concentrated in apps that rely on frequent updates, strict security, or changing APIs, while content-focused apps tend to adopt CI less.
These results encourage developers to adopt CI, especially in apps that require frequent updates, high reliability, or strong integration, indicating where CI practices may be most useful. 
To mitigate concerns about non-representative samples, we exclude categories with fewer than five apps, reducing the dataset from $n_{\text{non-CI}}=69$, $n_{\text{CI}}=177$ to $n_{\text{non-CI}}=61$, $n_{\text{CI}}=153$.

\begin{table}[!htbp]
\caption{Google Play Store Category Breakdown (sorted by total repositories): CI vs.\ Non-CI. Categories with the highest CI shares are marked \textit{bold}.}
\vspace{-7pt}
\label{tab:gps_category_breakdown_small}
\centering
\small  
{%
\begin{tabular}{@{} p{2.55cm} r r r r @{}}
\toprule
\textbf{Category} & \textbf{CI} & \textbf{Non-CI} & \textbf{Total} & \textbf{CI\%} \\
\midrule
TOOLS                  & 64 (26.0\%) & 32 (13.0\%) & 96 (39.0\%) & 66.7 \\
PRODUCTIVITY           & 21 (8.5\%)  & 12 (4.9\%)  & 33 (13.4\%) & 63.6 \\
GAME                   & 10 (4.1\%)  & 3 (1.2\%)   & 13 (5.3\%)  & 76.9 \\
MUSIC\_AND\_AUDIO      & 9 (3.7\%)   & 4 (1.6\%)   & 13 (5.3\%)  & 69.2 \\
COMMUNICATION          & 10 (4.1\%)  & 2 (0.8\%)   & 12 (4.9\%)  & 83.3 \\
\textbf{FINANCE}       & \textbf{10 (4.1\%)} & \textbf{0 (0.0\%)} & \textbf{10 (4.1\%)} & \textbf{100.0} \\
ENTERTAINMENT          & 6 (2.4\%)   & 2 (0.8\%)   & 8 (3.3\%)   & 75.0 \\
PERSONALIZATION        & 7 (2.8\%)   & 1 (0.4\%)   & 8 (3.3\%)   & 87.5 \\
\textbf{SOCIAL}        & \textbf{6 (2.4\%)} & \textbf{0 (0.0\%)} & \textbf{6 (2.4\%)} & \textbf{100.0} \\
PHOTOGRAPHY            & 2 (0.8\%)   & 3 (1.2\%)   & 5 (2.0\%)   & 40.0 \\
\textbf{MAPS\_AND\_NAV} & \textbf{5 (2.0\%)} & \textbf{0 (0.0\%)} & \textbf{5 (2.0\%)} & \textbf{100.0} \\
HEALTH\_AND\_FITNESS   & 3 (1.2\%)   & 2 (0.8\%)   & 5 (2.0\%)   & 60.0 \\
\midrule
\textbf{TOTAL}         & \textbf{153 (71.5\%)} & \textbf{61 (28.5\%)} & \textbf{214 (100\%)} & \textbf{76.9} \\
\bottomrule
\end{tabular}}
\Description{Google Play Store categories sorted by total repositories; categories with 100\% CI adoption are bolded.}
\vspace{-7pt}
\end{table}

\begin{rqbox}
\textbf{RQ1 Summary:}
Among repositories active in the 90 days before October~1,~2025, CI adopters are larger, more active, and release more frequently. For Google Play Store apps, CI correlates with substantially higher user engagement, while listing likelihood is similar for adopters and non-adopters. CI adoption is particularly common in utility- and integration-heavy categories, highlighting where frequent updates and reliability matter most.
\end{rqbox}

\subsection{RQ2: What are the distinct patterns of CI adopters?}
\subsubsection{\textbf{Motivation}}
Building on the RQ1 findings that CI adopters tend to be larger, more active, and more frequently updated projects, this research question explores \emph{how} CI adopters differ internally. Understanding the specific patterns and configurations within CI-adopting projects helps explain variations in development efficiency (e.g., faster releasing or quicker issue resolution) and reveals which CI practices or project characteristics are associated with more effective software delivery.

\subsubsection{\textbf{Approach}}
We build on the RQ1 dataset to perform a finer-grained analysis of CI adopters. As summarized in Table~\ref{tab:repo_metrics}, we extend the feature set by (i) identifying \emph{labeled} issues (rather than all issues) and (ii) normalizing commit, pull request, and project lifetime counts to a \emph{monthly} scale. We also detect the presence of common CI services, including Travis CI\footnote{\url{https://www.travis-ci.com/}}, CircleCI\footnote{\url{https://circleci.com/}}, GitLab CI\footnote{\url{https://about.gitlab.com/}}, and GitHub Actions\footnote{\url{https://docs.github.com/en/actions}}.

\begin{table}[t]
\caption{Repository-level metrics used in our analyses.}
\vspace{-7pt}
\centering
\small
\setlength{\tabcolsep}{6pt}
\renewcommand{\arraystretch}{1}

\begin{tabular}{p{0.42\linewidth} c p{0.46\linewidth}}
\hline
\textbf{Repository-level metric} & \textbf{DT*} & \textbf{Description} \\
\hline
Source\_Lines & N & Source lines of code for the repository. \\
no\_Contributors & N & Number of unique contributors who committed. \\
Project\_lifetime (months) & N & Age of the project in months (project creation date to end of observation). \\
(Average \# of pull requests) / month & N & Mean number of pull requests opened per month (PRs/month). \\
(Average \# of commits) / month & N & Mean number of commits per month (commits/month). \\
\# of CI services & N & Count of distinct CI services configured in the repository. \\
GitHub\_Actions & C & Whether GitHub Actions is configured/used in the repository. \\
Travis & C & Whether Travis CI is configured/used in the repository. \\
CircleCI & C & Whether CircleCI is configured/used in the repository. \\
GitLab & C & Whether GitLab CI is configured/used in the repository. \\
\hline
\end{tabular}

\vspace{4pt}
\footnotesize
\textit{DT*: (C) Categorical; (N) Numeric.}

\label{tab:repo_metrics}
\vspace{-10pt}
\end{table}

\vspace{3pt}
\noindent\textbf{\textit{Step 1: Data Collection.} }
Following Santos et al.~\cite{10.1145/3544902.3546244}, we identify bug-related issues as those labeled with any of \{\texttt{defect},
\texttt{error},
\texttt{bug},
\texttt{issue},
\texttt{mistake},
\texttt{incorrect},
\texttt{fault},
\texttt{flaw},
\texttt{crash},
\texttt{regression}\}. Repositories without such labels are excluded to ensure reliable defect-related metrics. The resulting dataset contains 442 CI-adopting repositories with at least one bug-labeled issue.

\noindent\textbf{\textit{Step 2: Data Preprocessing.}}
For each repository \(r\), we define its active duration in months as:
\[
M_r=\max\!\left(\varepsilon,\ \frac{\text{last\_activity\_date}_r-\text{first\_activity\_date}_r}{30.44}\right),
\]
where \(30.44 \approx 365.25/12\) converts days to average months, smoothing over month length variation. A small constant \(\varepsilon>0\) (e.g., 0.1) prevents division by zero for very short histories. For each count metric \(X \in \{\#\text{commits},\ \#\text{PRs opened},\ \#\text{PRs merged}\}\),
\[
X^{\text{per\_month}}_r=\frac{X^{\text{total}}_r}{M_r}.
\]

\noindent\textbf{\textit{Step 3: Quadrant-Based Analysis.}}
We exclude repositories with zero labeled bug issues, and compute bug density as:
\[
\text{bug\_density} \;=\; \frac{\texttt{bug\_issues\_labeled}}{\texttt{Source\_Lines}/1000}\, .
\]
Repositories are then split by the medians of tag lifetime (months) and bug density to form four groups:
\emph{fast\_low\_bugs},
\emph{slow\_low\_bugs},
\emph{fast\_high\_bugs}, and
\emph{slow\_high\_bugs}, which we visualize in a quadrant plot to highlight contrasting CI adoption characteristics.
The global medians are compute as follows:  
\[
\text{tag-release lifetime} = 15.5 \text{ days}, \quad 
\text{bug density} = 1.18 \text{ issues/KLoC}.
\]  

\noindent\textbf{\textit{Step 4: Modeling and Feature Importance.}}  
We train a multi-class \textit{Random Forest} classifier~\cite{breiman2001random} on the filtered dataset ($N=442$) to identify the project characteristics that best distinguish quadrant membership. 
The model uses 10 features capturing project scale, activity, CI service usage, and defect pressure: \texttt{Source\_Lines}, \texttt{no\_Contributors}, \texttt{Project\_lifetime (months)}, \texttt{(Avg.\# PRs)/ month}, \texttt{(Avg.\# commits)/month}, \texttt{\# CI services}, \texttt{GitHub\_Actions}, \texttt{Travis}, \texttt{CircleCI}, and \texttt{GitLab}.
Binary columns are encoded as integers (e.g., \texttt{GitLab}$\in\{0,1\}$).  
We apply a stratified 89.8\%/10.2\% train/test split ($397/45$) and fit the model with 500 trees, $\textit{max\_depth}=15$, $\textit{min\_samples\_split}=5$, and $\textit{min\_samples\_leaf}=2$.  
After training the Random Forest model, we apply permutation feature importance\footnote{\url{https://scikit-learn.org/stable/modules/permutation_importance.html}} to identify the most important features for each quadrant. The top features are then aggregated and ranked using \textsc{SK-ESD} clustering~\cite{tantithamthavorn2016empirical}, which provides a consolidated view of the key discriminators for quadrant membership.
We should note that our data is nearly evenly distributed across the four classes (approximately 110 projects per class and 10 features used for modeling). Therefore, as proven by Peduzzi et al~\cite{Peduzzi1996EPV}, an events-per-variable (EPV) ratio of 10 or greater is unlikely to cause overfitting.

\subsubsection{\textbf{Findings:}} We present detailed results for CI adopters, reporting the median \emph{tag release lifetime} and labeled \emph{bug-issue density} in each quadrant. 

\begin{figure}[!htbp]
  \vspace{-4pt}
  \centering
  \includegraphics[width=\linewidth]{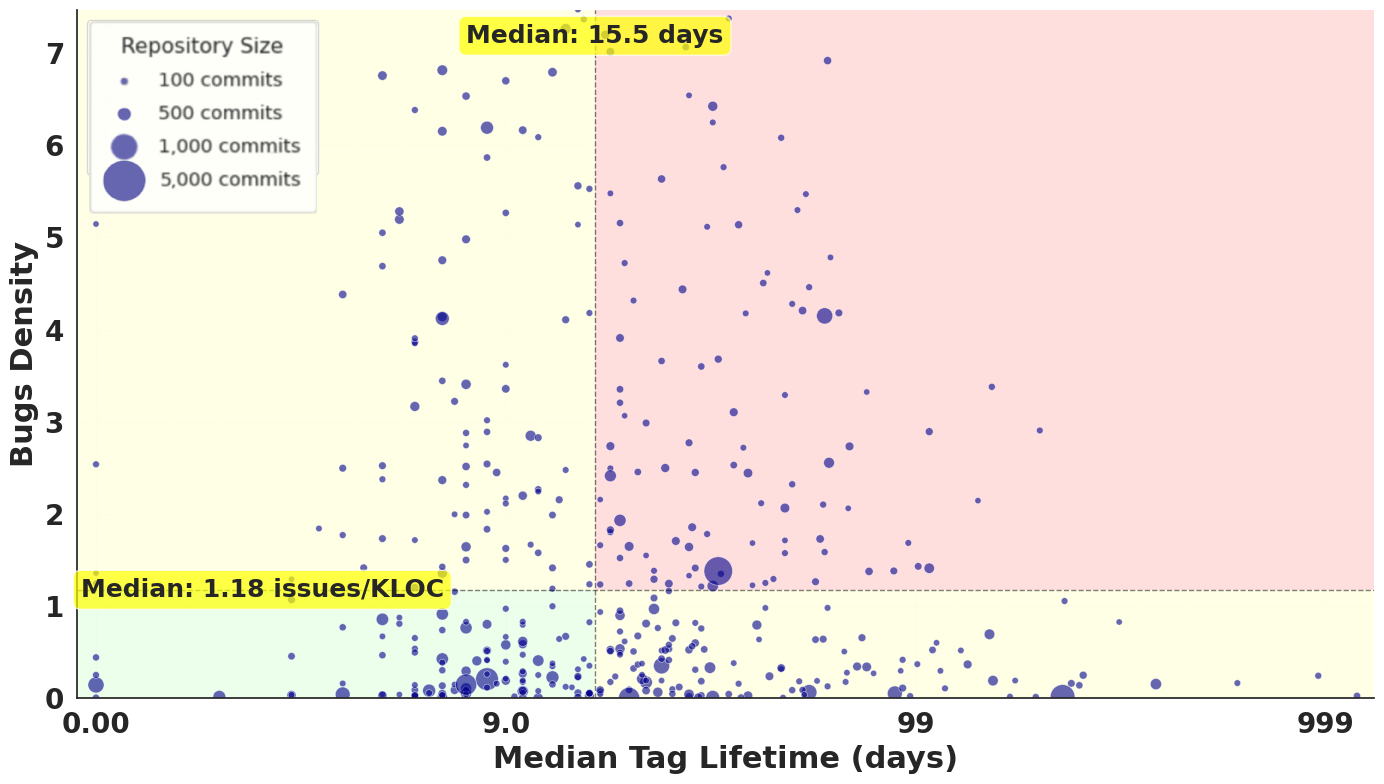}
  \vspace{-15pt}
  \caption{Quadrant plot for CI projects.}
  \Description{Quadrant plot for CI-adopting repositories. The figure shows four quadrants: the best quadrant (fast releasing with low bug density), the worst quadrant (slow releasing with high bug density), and the trade-off speed and defect density quadrants.}
  \label{fig:quadrant-analysis-summary-cicd}
  \vspace{-6pt}
\end{figure}

\vspace{3pt}
\noindent\textbf{Observation 2.1: Fast releasing do \emph{not} increase bug density among CI adopters.}
Splitting the \num{442} CI-adopting repositories by tag-release lifetime and bug density medians yields four nearly equal groups: Fast/Low = 108, Slow/Low = 113, Fast/High = 113, Slow/High = 108 (Figure~\ref{fig:quadrant-analysis-summary-cicd}).  
Frequent releases supported by CI do not compromise code quality, suggesting that disciplined CI pipelines enable rapid delivery without increasing defects.

The Best quadrant (Fast/Low) shows a median bug density of \(\num{0.23}\) issues/KLoC and a tag-release lifetime of \(\num{8.0}\) days, whereas the Worst quadrant (Slow/High) reaches \(\num{2.73}\) issues/KLoC and \(\num{32.0}\) days (Mann–Whitney \(U=0\), \(p<10^{-6}\) for both metrics). Projects in the Best quadrant also sustain strong development activity (median commits \(=\num{1145}\)), while other quadrants typically trade off speed or defect density.  
This indicates that maintaining fast releasing through CI does not inherently compromise software quality, as disciplined CI practices can support rapid, reliable delivery without increasing bug density.

\vspace{3pt}
\noindent\textbf{Observation 2.2: Project lifetime and scale/activity primarily discriminate CI adoption patterns, while CI service adds modest marginal signal.}
As Table~\ref{tab:sk_esd_ranking} depicts, \texttt{Project\_lifetime (months)} consistently ranks as the top discriminator across all four quadrants. Measures of project size and throughput: \texttt{Source\_Lines}, \texttt{no\_Contributors}, and monthly activity (\texttt{(Avg \# commits)/month}, \texttt{(Avg \# PRs)/month}), are the Top–5 features (Table~\ref{tab:top5_quadrant_block}). Overall, CI service indicators (\texttt{GitHub\_Actions}, \texttt{CircleCI}, \texttt{GitLab}) contribute little on average.  
In addition, the type of CI service can still provide informative discrimination. For example, in the \texttt{slow\_low\_bugs} quadrant, \texttt{Travis} and the number of CI services appear in the Top-5, and \texttt{GitHub\_Actions} also ranks highly.
This suggests that while project longevity and throughput are the main factors differentiating quadrants, specific CI services may help developers optimize maintenance-oriented or slow-moving projects.

{
\sisetup{
  scientific-notation=fixed, fixed-exponent=0,
  round-mode=places, round-precision=3,
  group-digits=integer, group-minimum-digits=4
}
\begin{table}[ht]
\centering
\caption{SK--ESD feature ranking (mean importance and dispersion).}
\vspace{-7pt}
\label{tab:sk_esd_ranking}
\begingroup
\small  
\begin{tabular}{@{} p{4.65cm}
                S[table-format=+1.3]
                S[table-format=1.3]
                S[table-format=2.0, round-mode=places, round-precision=0] @{}}
\toprule
\textbf{Feature} & {\textbf{Importance}} & {\textbf{Std}} & {\textbf{Rank}} \\
\midrule
\texttt{Project\_lifetime (months)}  & 0.3400 & 0.0580 & 1 \\
\texttt{Source\_Lines}               & 0.1044 & 0.0438 & 2 \\
\texttt{(Average \# of pull requests)/month} & 0.0704 & 0.0385 & 3 \\
\texttt{no\_Contributors}            & 0.0667 & 0.0448 & 4 \\
\texttt{GitHub\_Actions}             & 0.0000 & 0.0000 & 5 \\
\texttt{CircleCI}                    & 0.0000 & 0.0000 & 6 \\
\texttt{GitLab}                      & 0.0000 & 0.0000 & 7 \\
\texttt{\# of CI services}           & 0.0000 & 0.0081 & 8 \\
\texttt{(Average \# of commits)/month} & -0.0170 & 0.0302 & 9 \\
\texttt{Travis}                      & -0.0178 & 0.0089 & 10 \\
\bottomrule
\end{tabular}
\par\endgroup

\vspace{0.25em}
\footnotesize
\emph{Note:} Importance is the SK--ESD average across clusters; Std is the across-cluster dispersion.
Zeros indicate near-zero mean importance under this specification.
\vspace{-5pt}
\end{table}
}

\begin{table}[ht]
\centering
\caption{Top 5 features by quadrant in Random Forest.}
\vspace{-7pt}
\label{tab:top5_quadrant_block}
\begingroup
\setlength{\tabcolsep}{6pt}
\resizebox{0.99\linewidth}{!}{%
\begin{tabular}{@{} l l S[table-format=1.4] S[table-format=2.0] @{}}
\toprule
& \textbf{Metric} & {\textbf{Imp.}} & {\textbf{Rank}} \\
\midrule
\multirow{5}{*}{\textbf{FAST\_HIGH\_BUGS}} 
  & \texttt{Project\_lifetime (months)}              & 1.1340 & 1 \\
  & \texttt{Source\_Lines}                           & 0.3143 & 2 \\
  & \texttt{(Average \# of commits) / month}         & 0.3081 & 3 \\
  & \texttt{Travis}                                  & 0.2939 & 4 \\
  & \texttt{\# of CI services}                       & 0.2913 & 5 \\
\midrule
\multirow{5}{*}{\textbf{FAST\_LOW\_BUGS}} 
  & \texttt{Project\_lifetime (months)}              & 1.2994 & 1 \\
  & \texttt{no\_Contributors}                        & 0.2961 & 2 \\
  & \texttt{Source\_Lines}                           & 0.2751 & 3 \\
  & \texttt{\# of CI services}                       & 0.2417 & 4 \\
  & \texttt{(Average \# of commits) / month}         & 0.2260 & 5 \\
\midrule
\multirow{5}{*}{\textbf{SLOW\_HIGH\_BUGS}} 
  & \texttt{Project\_lifetime (months)}              & 1.1543 & 1 \\
  & \texttt{(Average \# of commits) / month}         & 0.3024 & 2 \\
  & \texttt{Source\_Lines}                           & 0.2235 & 3 \\
  & \texttt{Travis}                                  & 0.2233 & 4 \\
  & \texttt{GitHub\_Actions}                         & 0.2021 & 5 \\
\midrule
\multirow{5}{*}{\textbf{SLOW\_LOW\_BUGS}} 
  & \texttt{Project\_lifetime (months)}              & 1.2653 & 1 \\
  & \texttt{Travis}                                  & 0.7582 & 2 \\
  & \texttt{\# of CI services}                       & 0.5684 & 3 \\
  & \texttt{GitHub\_Actions}                         & 0.5426 & 4 \\
  & \texttt{GitLab}                                  & 0.3306 & 5 \\
\bottomrule
\end{tabular}}
\par\endgroup
\vspace{-10pt}
\end{table}

\vspace{-7pt}
\begin{rqbox}
\textbf{RQ2 Summary:}
Fast release cycles do not inherently increase defect density. Quadrant analysis reveals that CI enables fast releasing without sacrificing quality. Project lifetime and bug density are the key factors differentiating project performance, while CI service usage mainly distinguishes slower, more stable projects. Overall, CI adoption can help maintain a disciplined release cycle and low defect load.
\end{rqbox}

\subsection{RQ3: What are the evolution characteristics of CI adoption?}
\subsubsection{\textbf{Motivation}}
Studying the impact of CI adoption is crucial for understanding how it affects development activity and user outcomes. While CI is often associated with improved efficiency, its effects can vary across repositories. By analyzing changes in commits, pull requests, and merge times, along with user-facing outcomes like reviews and ratings, we aim to identify which projects benefit most from CI adoption and whether it improves the user experience. This can help developers optimize their CI practices for better results.

\subsubsection{\textbf{Approach:}}
We collect and analyze repository-level metrics for this research question. Building on the RQ2 setup, we focus on repositories active in the 90 days around CI adoption and also examine each project’s full lifetime. For code-hosting outcomes, we compute before/after changes in (i) commits, (ii) pull requests opened, (iii) pull requests merged, and (iv) time to merge. We detect CI adoption from workflow-configuration history and align it with each repository’s activity timeline. Data are obtained via the GitHub REST API; user-facing app outcomes (monthly review volume and average star rating) are retrieved from the Google Play Store using \texttt{google\_play\_scraper}\footnote{\url{https://pypi.org/project/google-play-scraper}}.

\vspace{3pt}
\noindent\textbf{\textit{Step 1: Data Collection.}}
Focusing on repositories active in the 90-day window around CI adoption, we collect (i) commits, (ii) pull requests opened, (iii) pull requests merged, and (iv) time to merge from the GitHub REST API. For user-facing outcomes, we query the Google Play Store via \texttt{google\_play\_scraper} and obtain review histories, from which we compute total review counts and average star ratings over the pre- and post-adoption windows. 

\vspace{3pt}
\noindent\textbf{\textit{Step 2: data Preprocessing.}}
Some repositories adopted CI at project inception and thus lack a pre-adoption window; others had no early pull-request activity. Consequently, analyses are metric-specific (pairwise inclusion). From the original 177 repositories, the usable samples are: commits \(n=172\), PRs opened \(n=137\), PRs merged \(n=133\), and time to merge \(n=127\).
For Google Play Store outcomes, some linked apps lack a valid pre-adoption window (the app launched after adoption or had no reviews/ratings recorded in the pre window). Thus, Play-based analyses are also metric-specific: review-volume rate ratio \(n=97\) and rating-change rate ratio \(n=95\).

\vspace{3pt}
\noindent\textbf{\textit{Step 3: Data Analysis.}}
We perform paired, within-repository comparisons using the full project history before vs.\ after CI adoption. The pre-adoption window spans from the first commit/PR to the CI adoption date; the post-adoption window spans from the adoption date to the last commit/PR. Metrics are month-normalized by dividing activity counts by the duration of each window in months (computed as $\text{days}/30.44$). For each repository $r$, we report four rate ratios:

\begin{enumerate}
\item 
\noindent\textit{Commits rate ratio.}
Let $C^{\text{pre}}_r$ and $C^{\text{post}}_r$ be the commit counts in the pre- and post-adoption windows. Let $M^{\text{pre}}_r$ and $M^{\text{post}}_r$ be the window lengths in months. Define:
\[
\text{commits\_increase\_rate}_r
=\frac{C^{\text{post}}_r / M^{\text{post}}_r}
       {C^{\text{pre}}_r  / M^{\text{pre}}_r}.
\]
Values $>1$ indicate higher commit throughput after adoption.

\smallskip
\noindent\textit{PRs-opened rate ratio.}
Let $P^{\text{pre}}_r$ and $P^{\text{post}}_r$ be the numbers of pull requests opened in each window. Then:
\[
\text{prs\_increase\_rate}_r
=\frac{P^{\text{post}}_r / M^{\text{post}}_r}
       {P^{\text{pre}}_r  / M^{\text{pre}}_r}.
\]
This captures the change in PR creation activity.

\smallskip
\noindent\textit{Merged-PRs rate ratio.}
Let $N^{\text{pre}}_r$ and $N^{\text{post}}_r$ be merged PR counts per window. We compute:
\[
\text{merged\_prs\_increase\_rate}_r
=\frac{N^{\text{post}}_r / M^{\text{post}}_r}
       {N^{\text{pre}}_r  / M^{\text{pre}}_r}.
\]
Larger values reflect more merges per month after adoption.

\item
\noindent\textit{Merge-time speed-up.}
For merged PRs, define merge time as $\texttt{merged\_at}-\texttt{created\_at}$ (days). Let $\widetilde{T}^{\text{pre}}_r$ and $\widetilde{T}^{\text{post}}_r$ be the median merge times in the two windows. We report the inverse ratio (speed-up):
\[
\text{merge\_time\_increase\_rate}_r
=\frac{\widetilde{T}^{\text{post}}_r}{\widetilde{T}^{\text{pre}}_r},
\]
so that values $<1$ indicate faster merging after adoption.

\end{enumerate}

\smallskip
\noindent\textbf{\textit{Interpretation.}}
Rates $>1.0$ indicate an increase after CI; $=1.0$ indicates no change; $<1.0$ indicates a decrease. Ratios with undefined denominators are treated as missing and excluded, as no data before CI.

\subsubsection {\textbf{Findings:}} We present detailed before vs. after results around CI adoption.

\begin{figure}[t]
  \centering
  \includegraphics[width=\linewidth]{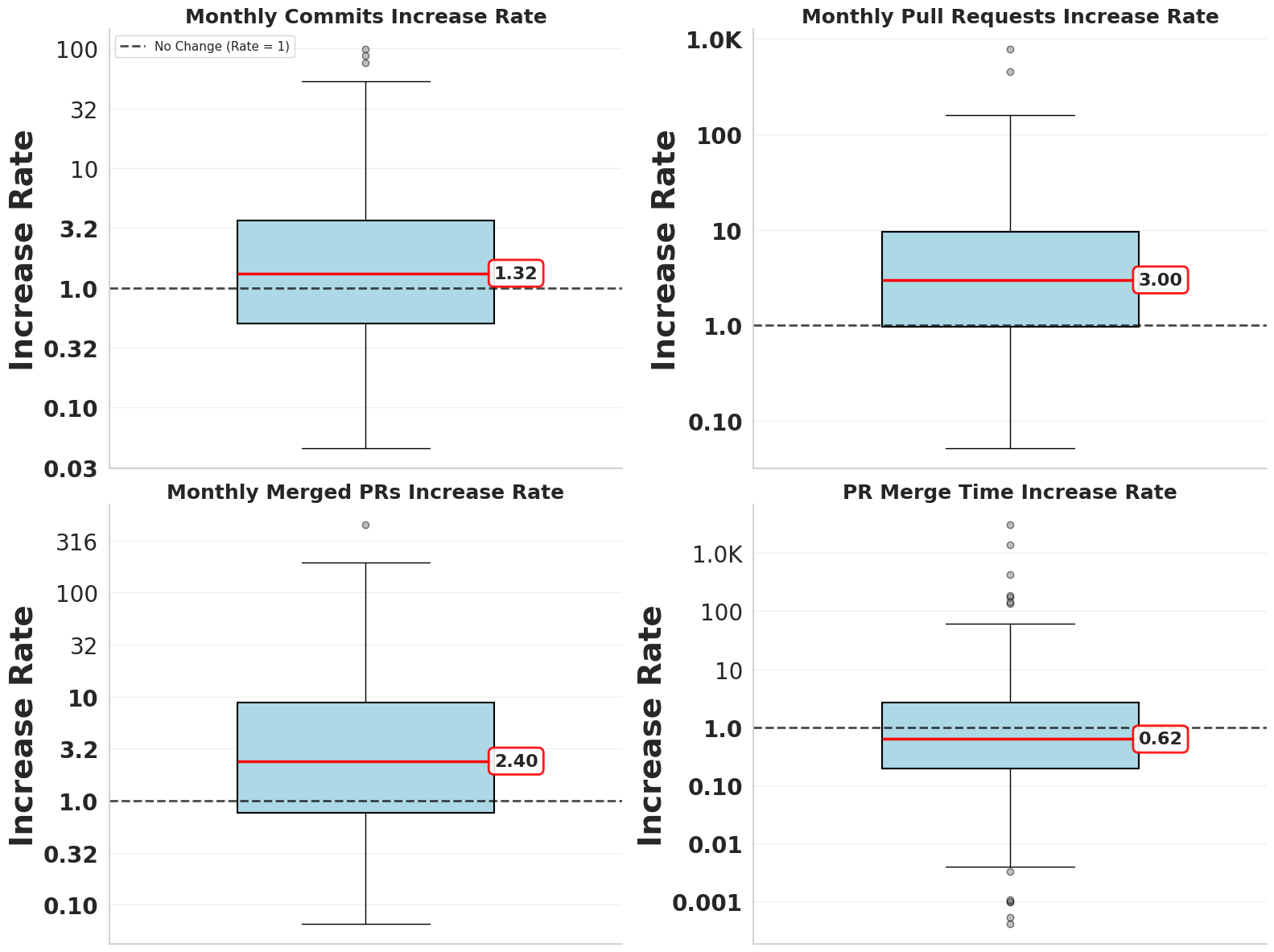}
  \caption{Before vs. after rate ratios around CI adoption.}
  \Description{Monthly commits increase rate, monthly pull requests (PR) increase rate, monthly merged PR increase rate, and PR merge time increase rate for CI-adopters.}
  \label{fig:before_after_repos}
  \vspace{-5pt}
\end{figure}

\vspace{2pt}
\noindent\textbf{Observation 3.1: Development activity (in terms of commits and PRs) improves after CI adoption.}
After adopting CI, development performance improves, with repositories seeing increases in commits, pull requests (PRs) opened, and PRs merged. However, the introduction of more rigorous quality gates and increased PR concurrency leads to some trade-offs in merge speed.
As shown in Figure~\ref{fig:before_after_repos} and Table~\ref{tab:before_after_rate_ratios}, development activity generally increases post-CI adoption: 59.3\% of repositories see more commits, 74.5\% see more PRs opened, and 72.2\% see more PRs merged. The median merge-time speed-up is 1.60, though only 41.4\% of repositories merge faster after CI adoption, indicating that additional quality controls (such as status checks and required reviews) introduce some latency. The distributions of these metrics show heavy right tails, with means far exceeding medians, suggesting that while most repositories show moderate improvements, a few repositories experience substantial post-adoption surges (e.g., PRs opened up to 777 and merge-time ratio up to 3065.52).
Therefore, developers should be prepared for trade-offs in merge speed due to added quality assurance measures. A balance between speed and rigor is recommended, especially for large projects with higher PR volumes, as the degree of improvement can vary significantly across projects.

\begin{table}[t]
\caption{Before vs.\ After CI Adoption: 
Values $>1$ indicate an increase after adoption; for merge time, the ratio is reported as a speed-up (before/after).}
\vspace{-7pt}
\label{tab:before_after_rate_ratios}
\centering
\resizebox{\columnwidth}{!}{%
\begin{tabular}{@{} l r r r r r l @{}}
\toprule
\textbf{Metric} & \textbf{Count} & \textbf{Median} & \textbf{Mean} & \textbf{Min} & \textbf{Max} & \textbf{Up / Down} \\
\midrule
Commits           & \num{172} & \num{1.32} & \num{5.39}  & \num{0.05}   & \num{98.94}   & \num{102} (59.3\%) / \num{70} (40.7\%) \\
PRs Opened        & \num{137} & \num{3.00} & \num{20.02} & \num{0.00}   & \num{777.00}  & \num{102} (74.5\%) / \num{35} (25.5\%) \\
PRs Merged         & \num{133} & \num{2.40} & \num{13.67} & \num{0.00}   & \num{447.00}  & \num{96}  (72.2\%) / \num{37} (27.8\%) \\
Merge Time      & \num{127} & \num{1.60} & \num{46.03} & \num{0.00}   & \num{3065.52} &  \num{75} (58.6\%) /\num{53}  (41.4\%)  \\
\bottomrule   
\end{tabular}}
\Description{Summary statistics for four before vs. after rate ratios.}
\vspace{-10pt}
\end{table}

\begin{figure}[!htbp]
  \centering
  \includegraphics[width=\linewidth]{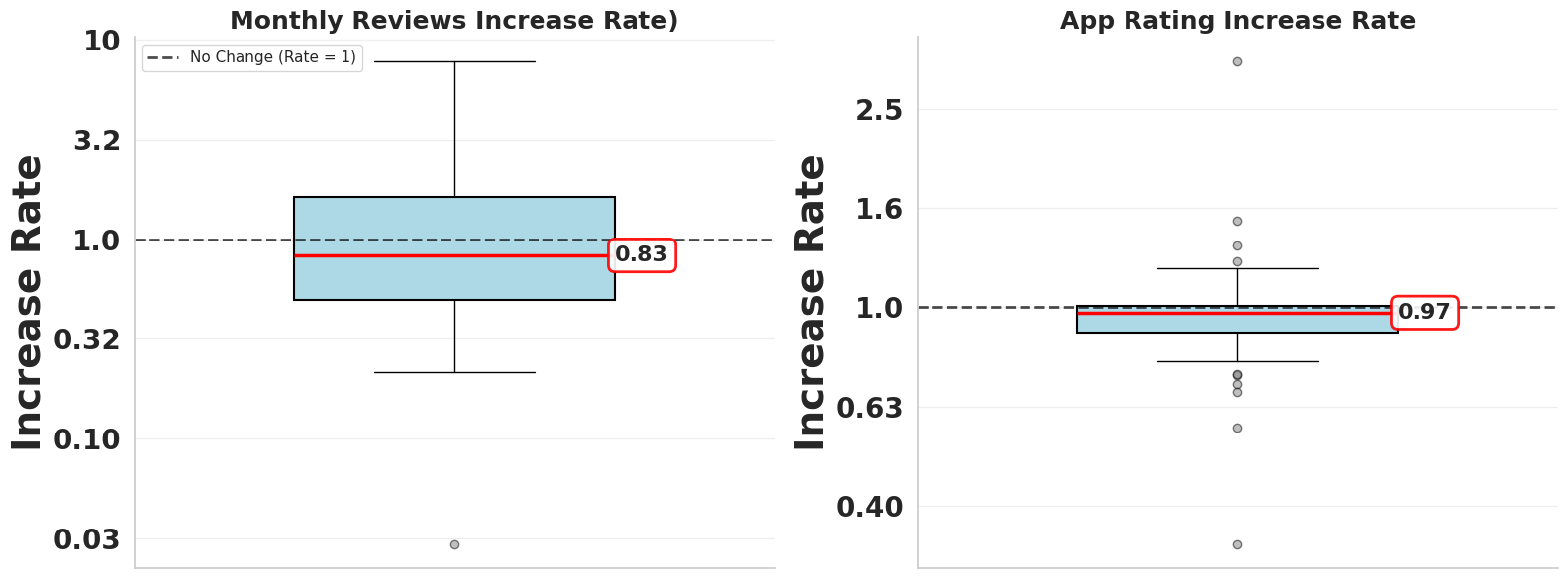}
  \caption{Google Play Store Outcomes Before vs. After CI Adoption.}
  \Description{Monthly reviews increase rate and app rating increase rate for CI-adopters.}
  \label{fig:before_after_gps}
  \vspace{-10pt}
\end{figure}

\vspace{3pt}
\noindent\textbf{Observation 3.2: No systematic improvement in user-facing outcomes after CI adoption.}
CI adoption does not appear to lead to systematic improvements in user-facing outcomes, such as Google Play Store reviews and star ratings, in the short run.As shown in Figure~\ref{fig:before_after_gps} and Table~\ref{tab:play_before_after_rate_ratios}, there is no significant positive shift in median review volume or star ratings after CI adoption. In fact, the median review rate declines slightly to $0.83$, with fewer than half of apps experiencing increases in reviews (41.2\%). Similarly, median star ratings remain unchanged ($0.97$), with only 31.6\% of apps showing improvements. The distributions are right-skewed, with means exceeding medians: reviews ($1.21$ vs.\ $0.83$) and ratings ($0.97$ vs.\ $0.97$), indicating that a small subset of apps saw large gains, but these do not significantly affect the overall trend.
This suggests that developers should be cautious in expecting immediate or substantial improvements in user-facing outcomes, such as app ratings or review volume. Other factors, including app content, user experience, and marketing efforts, likely play a larger role in shaping user feedback on platforms like the Google Play Store.

\begin{table}[ht]
\caption{Google Play Store Outcomes Before vs.\ After CI Adoption: Rate Ratio Summary. Values $>1$ indicate an increase after adoption.}
\vspace{-7pt}
\label{tab:play_before_after_rate_ratios}
\centering
\resizebox{\columnwidth}{!}{%
\begin{tabular}{@{} l r r r r r l @{}}
\toprule
\textbf{Metric} & \textbf{Count} & \textbf{Median} & \textbf{Mean} & \textbf{Min} & \textbf{Max} & \textbf{Up / Down} \\
\midrule
Monthly Reviews  & \num{97} & \num{0.83} & \num{1.21} & \num{0.00} & \num{7.83} & \num{40} (41.2\%) / \num{57} (58.8\%) \\
Average Rating   & \num{95} & \num{0.97} & \num{0.97} & \num{0.33} & \num{3.13} & \num{30} (31.6\%) / \num{64} (67.4\%) \\
\bottomrule
\end{tabular}}
\Description{Summary statistics for Google Play Store rate ratios (monthly reviews and average rating) computed over symmetric 90-day windows.}
\end{table}

\begin{rqbox}
\textbf{RQ3 Summary:}
CI adoption likely boosts development activity, with more commits and pull requests opened and merged, though median merge times may lengthen due to stricter pipelines. In contrast, user-facing metrics such as app reviews and ratings show no consistent short-term improvement, suggesting that CI primarily enhances internal development processes rather than immediate user outcomes.
\end{rqbox}

\subsection{RQ4: What explains the variation in before- and after-adoption performance?}
\subsubsection{\textbf{Motivation}}
In RQ3, we find significant variability in the association of CI adoption with commit activity, pull requests, and merged PRs across different projects. To better understand which repositories benefit the most from CI adoption, it is important to examine the factors driving this variability. By conducting a pre/post analysis around each repository’s CI adoption date, we can estimate relative changes in activity metrics and explore how these changes differ across adoption-lifetime quartiles, app categories, repository scale, and CI service configurations. This helps pinpoint key conditions associated with CI's effectiveness, offering practical insights into how developers can maximize the benefits of continuous integration.

\subsubsection{\textbf{Approach:}}
Building upon the filtered dataset from RQ3, which consists of 177 active repositories, we conduct a temporal analysis to examine the CI adoption lifecycle. To quantify the duration of CI usage across repositories, we establish a time window by calculating the period between the CI adoption date and the repository's last commit activity.

\vspace{3pt}
\noindent\textbf{\textit{Step 1: Data Preprocessing.}}
Using the same data and preprocessing as in RQ3 (i.e., $n = 177$), we remove repositories with missing merge times, resulting in 127 repositories.
We then standardize all variables using StandardScaler\footnote{\url{https://scikit-learn.org/stable/modules/generated/sklearn.preprocessing.StandardScaler.html}} to ensure that each feature contributes equally to distance-based analyses and to remove scale effects, which is important when comparing metrics with different units or ranges.

\vspace{3pt}
\noindent\textbf{\textit{Step 2: Data Analysis.}}
\noindent We compute \emph{Adoption Lifetime} for each repository as the time between the CI creation date and the most recent activity date:

\resizebox{0.92\linewidth}{!}{$
\mathrm{AdoptionLifetimeMonths}_i
= \frac{\mathrm{Last Activity Date}_i - \mathrm{CI Create Date}_i}{30.44}
$}

\vspace{4pt}
We then apply $k$-means clustering~\cite{mcqueen1967some} with a multi-dimensional feature space that captures changes in development and release activity after CI/CD adoption, 
including commit increase rate, pull request increase rate, merged pull requests increase rate, and release lifetime increase rate, to partition repositories into groups with similar characteristics.

\subsubsection{\textbf{Findings:}} We present detailed explanatory results on CI adoption

\vspace{3pt}
\noindent\textbf{Observation 4.1: CI adoption lifetimes exhibit substantial heterogeneity.} 
Repository lifetime shows considerable variation, with the median project age around 39.5 months (~3.3 years). This heterogeneity helps explain the observed differences in post-CI adoption performance, as newer projects can experience sharp gains, while longer-term projects may see smaller incremental improvements.
We observe that the median lifetime of repositories is 39.46 months (~3.29 years), with the mean slightly higher at 41.15 months (~3.43 years). The distribution is slightly right-skewed, with a range of approximately 131.87 months (from 1.45 months to 133.31 months) and an interquartile range (IQR) of about 36.33 months (~3.03 years). This wide variation helps explain the performance differences observed in RQ3: newer projects can see rapid performance improvements from a low baseline, while projects with longer histories may only show marginal gains due to already having optimized practices.
The long median repository age suggests that most projects have had sufficient time to realize CI benefits. However, developers should expect more significant performance gains from newer projects, while older projects may see more modest improvements due to prior optimizations.

\begin{figure}[t]
  \centering
  \vspace{-8pt}
  \includegraphics[width=\linewidth]{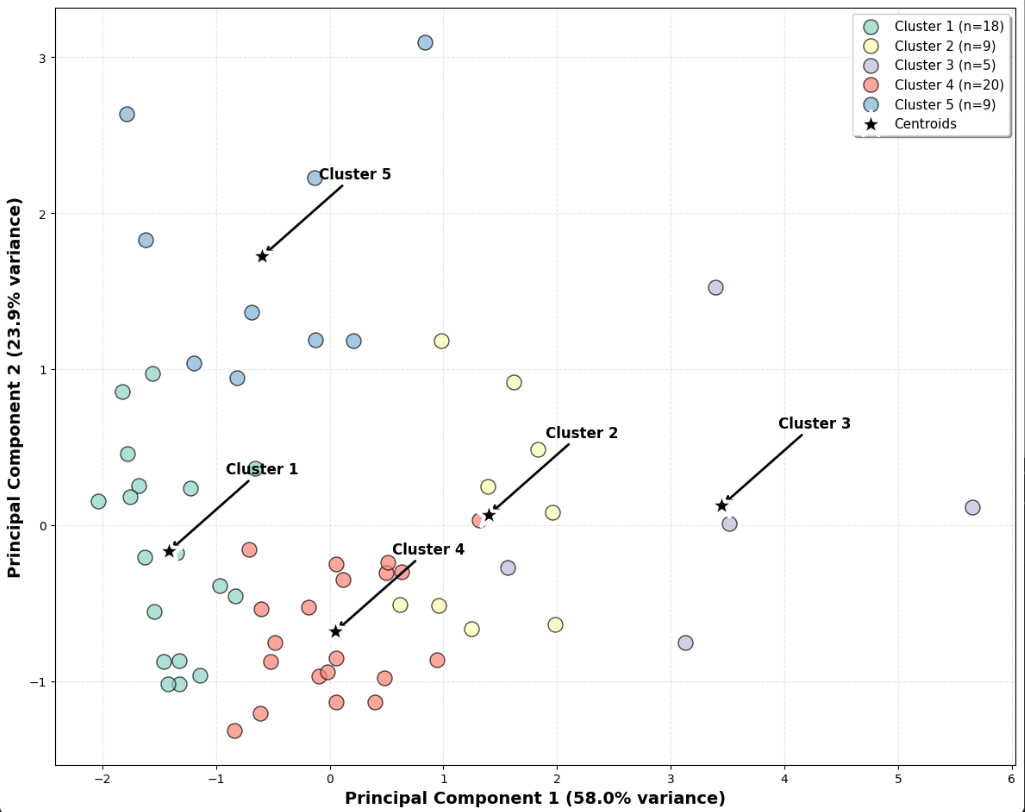}
  \vspace{-15pt}
  \caption{$k$-means clustering with $k=5$ on repositories (log-transformed features). Points are colored by cluster; stars mark centroids.}
  \Description{The identified clusters using $k$-means clustering with $k=5$ on repositories with log-transformed features. Points are colored by cluster; stars mark centroids.}
  \vspace{-7pt}
  \label{fig:kmeans_repos}
\end{figure}

\vspace{3pt}
\noindent\textbf{Observation 4.2: Post CI-adoption development patterns vary by release timing and activity.}
As shown in Figure~\ref{fig:kmeans_repos}, clustering the before/after rate ratios reveals three primary post-adoption patterns across 61 repositories, as follows. In this, all the 4 metrics must not be NAN.

\begin{itemize}
    \item \textbf{Cluster4} shows \textbf{increased} commits/PRs and \textbf{shorter} release lifetimes.
    \item \textbf{Cluster1} shows \textbf{decreased} commits/PRs with \textbf{longer} release lifetimes.
    \item \textbf{Cluster2}, \textbf{Cluster3}, \textbf{Cluster5} show \textbf{increased} activity with \textbf{longer} releases, but with substantial variability in release duration.
\end{itemize}

\vspace{4pt}
\noindent\textbf{Cluster interpretations:}
\begin{itemize}\itemsep0.25em
  \item \textbf{Cluster1 (n=18): Activity down, releases longer:} commits \( \downarrow\) (0.57), PRs opened \( \downarrow\) (0.53), PRs merged \( \downarrow\) (0.54), release lifetime lengthens (3.82).
  \item \textbf{Cluster2 (n=9): PR-heavy gains, releases longer:} commits \( \uparrow\) (1.26), PRs opened \( \uparrow\) (4.81), PRs merged \( \uparrow\) (3.63), release lifetime lengthens (2.28).
  \item \textbf{Cluster3 (n=5): Very large commit gains, releases longer:} commits \( \uparrow\) (15.60), PRs opened \( \uparrow\) (7.61), PRs merged \( \uparrow\) (7.92), release lifetime lengthens (2.30).
  \item \textbf{Cluster4 (n=20): Increases with shorter releases:} commits \( \uparrow\) (1.83), PRs opened \( \uparrow\) (1.43), PRs merged \( \uparrow\) (1.42), release lifetime shortens (0.93).
  \item \textbf{Cluster5 (n=9): Modest gains, releases extremely longer:} commits \( \uparrow\) (1.29), PRs opened \( \uparrow\) (1.67), PRs merged \( \uparrow\) (1.51), release lifetime \emph{much} longer (28.13).
\end{itemize}

\noindent This clustering reveals that Increased development activity does not always lead to faster releases. While Cluster4 (n=20) demonstrates a continuous delivery pattern with more commits/PRs and shorter release lifetimes, the combined groups of Cluster2, Cluster3, and Cluster5 (n=23) show that higher activity can coincide with longer release lifetimes. Notably, Cluster5 shows an extremely large release-lifetime ratio (28.13), which inflates the means and contributes to the wide dispersion observed in RQ3. This suggests that post-adoption pipelines may introduce additional quality gates, such as testing, reviews, and compliance checks, that intentionally extend release cycles to improve stability and quality. Developers should recognize that longer release lifetimes are not necessarily a sign of inefficiency, but rather a trade-off for enhanced rigor and quality in the release process.\vspace{4pt}

\begin{table}[ht]
\centering
\caption{Cluster characteristics (original scale): \emph{means} of before/after rate ratios. Release lifetime is \emph{after/before} ($<\!1$ = shorter).}
\vspace{-7pt}
\label{tab:cluster_char_orig}
\begingroup
\small
\setlength{\tabcolsep}{4pt}
\resizebox{\columnwidth}{!}{%
\begin{tabular}{@{} l r
    S[table-format=5.2]
    S[table-format=5.2]
    S[table-format=5.2]
    S[table-format=6.2]
    @{}}
\toprule
\textbf{Cluster} & \textbf{n} &
\multicolumn{1}{c}{\textbf{Commits}} &
\multicolumn{1}{c}{\textbf{PRs opened}} &
\multicolumn{1}{c}{\textbf{PRs merged}} &
\multicolumn{1}{c}{\textbf{Release lifetime (after/before)}} \\
\midrule
1 & 18 & 0.57 & 0.53 & 0.54 & 3.82 \\
2 &  9 & 1.26 & 4.81 & 3.63 & 2.28 \\
3 &  5 & 15.60 & 7.61 & 7.92 & 2.30 \\
4 & 20 & 1.83 & 1.43 & 1.42 & 0.93 \\
5 &  9 & 1.29 & 1.67 & 1.51 & 28.13 \\
\bottomrule
\end{tabular}}
\par\endgroup
\end{table}

\begin{rqbox}
\textbf{RQ4 summary:}
Three primary post-CI adoption patterns observed in our dataset: (1) increased activity with shorter release lifetimes, (2) decreased activity with longer release lifetimes, and (3) increased activity with longer release lifetimes. Interestingly, more development activity does not always lead to faster releases; longer lifetimes often reflect the addition of quality controls, such as testing and reviews, rather than inefficiency.
\end{rqbox}
\vspace{-7pt}

\section{Implications}
\label{sec:Implications}
This section discusses the practical and research implications of our findings.

\vspace{3pt}
\noindent\textbf{For developers.}  
CI adopters are observed to be larger and more active, with faster releasing and higher user engagement (RQ1). Among adopters, faster releasing is not associated with higher bug density, indicating that well-structured CI can maintain quality even at higher development speed (RQ2). After adoption, repositories increase development activity (commits +59\%, PRs +74\%), but median merge times prolong, and Google Play Store metrics remain largely unchanged (RQ3). Developers should therefore complement CI with release and feedback practices to improve user-facing outcomes. The three observed adoption trajectories (RQ4) suggest that CI effects vary depending on project practices and pipeline configurations, emphasizing the need to tailor CI setups to team and project goals.

\vspace{3pt}
\noindent\textbf{For researchers.}  
The diversity in adoption outcomes (RQ4) points to open questions about how project characteristics, pipeline configurations, and team practices shape CI effectiveness. Future work could investigate longitudinal CI adoption, cross-domain comparisons (e.g., Android vs. ML projects), and socio-technical factors that can influence productivity, release speed, and user engagement.

\section{Threats to Validity}
\label{sec:Threats_to_Validity}
\textbf{Construct validity.}
(i) \emph{CI adoption.} We infer adoption from repository workflow/configuration histories and release tags. Projects that enable CI via web dashboards or less common providers may be missed or have misdated adoption. 
(ii) \emph{Metric proxies.} Repository signals (e.g., commit frequency, issue counts, SLOC) are proxies for productivity and quality; they may not fully capture latent constructs (e.g., developer effort, code health).
In particular, using GitHub Issues as a proxy for bug density has limitations: though we search for bug-related keywords (e.g., “defect,” “bug,” “crash”), these heuristics are not fully robust and may miss relevant cases or include false positives. To detect CI adoption, we identify repositories that contain CI configuration (YAML) files. While this method confirms that developers have configured CI, it cannot verify whether the CI pipeline is actively executed in practice. Nevertheless, this approach is the standard and most practical proxy for CI adoption in large-scale repository mining studies~\cite{10.1145/3544902.3546244}.

\vspace{3pt}
\noindent\textbf{Internal validity.}
Unobserved confounders (team size, domain, or governance model) may influence both CI adoption and outcomes. Though we control for observable repository characteristics, residual confounding may remain. Our 90-day activity window reduces long-run drift but may omit slower processes.
We should note that our results do not imply causality; they rather summarize patterns and associations observed in the current data.

\vspace{3pt}
\noindent\textbf{External validity.}
Our dataset focuses on Android apps and excludes iOS. We expect core development dynamics (e.g., commit frequency) to be broadly comparable across mobile operating systems, but this remains an assumption; platform-specific effects may limit generalizability. The study is limited to Android apps because app details are publicly available and crawlable on the Google Play Store, unlike other platforms such as the Apple App Store. Future work should extend this analysis to additional platforms. Results may also differ for non-mobile or closed-source projects.

\section{Related Work}
\label{sec:Related_Work}
\subsection{Mobile app release practices}
Prior studies examined release behaviors and their effects on app quality and user satisfaction. 
For example, \citet{DBLP:journals/ese/McIlroyAH16} found that frequent releases are associated with higher Google Play ratings, while \citet{DBLP:conf/sigsoft/Dominguez-Alvarez19} observed that Android apps release more often than iOS apps. 
Other studies analyzed emergency releases~\cite{Bad_Updates_TSE,Emergency_Updates_EMSE,Islem_EMSE} and release communication through notes~\cite{DBLP:journals/ese/YangHZH22}, showing diverse developer practices and challenges in managing updates. 
\citet{DBLP:conf/wcre/NayebiAR16} and \citet{DBLP:conf/esem/NayebiFR17} surveyed developers and modeled release deployment likelihood, linking structured release processes with higher app success. 
Beyond mobile apps, ecosystem-level studies have examined how release activity evolves over project lifecycles and how sustained slowdowns or prolonged inactivity often precede project abandonment. \citet{11025680}. showed that declining release activity is a common signal of reduced maintenance in large software ecosystems, and that release frequency alone is insufficient to characterize project sustainability. Our work complements this line of research by connecting release behavior with CI adoption, offering empirical evidence on how CI supports faster, more regular, and higher-engagement releases.

\subsection{CI adoption and practices}
Several studies explored CI adoption and challenges.
\citet{DBLP:conf/kbse/HiltonTHMD16} found limited CI usage on GitHub due to developers’ lack of experience, and \citet{chopra2025multici} studied how projects gradually adopt and evolve multiple CI services over time.
Several studies~\cite{widder2019conceptual,zampetti2020empirical,abrokwah2025empirical} identified CI pain points such as configuration smells, tool inconsistencies, and high complexity and heterogeneity in CI configurations.
To address these challenges, recent work has examined automation and evolution in CI usage, including automatic generation of CI configurations~\cite{ghaleb2025llm4ci} and CI service migration~\cite{hossain2025cigrate}. \citet{bouzenia2024resource} and \citet{jin2021helped} highlighted CI’s resource costs and feedback trade-offs, while \citet{ghaleb2019empirical,ghaleb2019studying,ghaleb2023interplay} studied configuration patterns and their effects on build duration and failure. \citet{ghaleb2023interplay,ghaleb2019empirical} further showed that attempts to improve CI build duration or reliability often involve trade-offs, as common workarounds (e.g., retries or extended timeouts) can slow builds without guaranteeing success.
Our work instead quantifies CI’s broader associations with development activity, release speed, and user outcomes, focusing on open-source Android projects.

\subsection{CI for mobile apps}
CI research in mobile development remains limited. 
\citet{DBLP:conf/mobilesoft/PoleseHT22} and \citet{DBLP:journals/smr/LiuLLMG24} showed that many Android projects do not build reliably, indicating difficulties in maintaining stable pipelines. 
\citet{DBLP:conf/esem/WangZXY23} found that only 40\% of apps use CI, mostly for unit testing. 
Similar to our study, \citet{ghaleb2025android} investigated four CI services, including \textsf{GitHub~Actions}, \textsf{Travis~CI}, \textsf{CircleCI}, and \textsf{GitLab~CI}, on a curated set of 2.5K Android apps from \textsf{GitHub} and \textsf{F-Droid}. They uncovered lack of commonalities among these services and identified trade-offs when selecting certain CI services.
Our work complements theirs by analyzing the effects of CI adoption on development activity, release patterns, and user-facing outcomes, offering a more detailed view of CI usage in mobile apps.

\section{Conclusion}
\label{sec:Conclusion}
This study analyzed 2,542 open-source Android projects, linking CI adoption with development activity and Google Play Store outcomes through statistical, machine learning, and clustering analyses.
CI adoption aligns with project maturity and complexity rather than simplicity. Adopters are larger, more active, and concentrated in integration-heavy domains (e.g., Finance and Productivity), suggesting CI supports scaling and reliability needs.
Faster releasing does not raise defect rates: top projects pair short 8-day releases with low bug density (0.23 issues/KLoC). Project lifetime and bug density dominate predictive importance, indicating disciplined CI sustains speed and quality.
Post-adoption, activity rises (commits +59\%, PRs +75\%) but merges slow and user ratings remain stable, showing CI boosts development throughput but not necessarily delivery pace or user satisfaction.
CI improves engineering efficiency and process control, but its impact on external success depends on complementary release and feedback practices.

\vspace{3pt}
\noindent\textbf{Future work.}
We plan in future research to survey developers to understand motivations, perceived challenges, and organizational barriers in CI adoption. Future work should also compare and contrast CI adoption across domains such as Android, web, artificial intelligence, and machine learning projects, and analyze how evolving CI toolchains and automation depth influence delivery outcomes and developer experience.

\section*{Artifact Availability}

A replication package, including scripts, data, and raw results, used to produce the findings of this study, is available online at Figshare~\cite{our_replication_package}.

\begin{acks}
Funding: We acknowledge the support of the Natural Sciences and Engineering Research Council of Canada (NSERC): [RGPIN-2021-03969] and [RGPIN-2025-05897].
\end{acks}

\balance
\bibliographystyle{ACM-Reference-Format}
\bibliography{References}

@article{ghaleb2025android,
  title={CI/CD Configuration Practices in Open-Source Android Apps: An Empirical Study},
  author={Ghaleb, Taher and Abduljalil, Osamah and Hassan, Safwat},
  journal={ACM Transactions on Software Engineering and Methodology},
  year={2024},
  publisher={ACM New York, NY}
}

@inproceedings{10.1145/3544902.3546244,
author = {Santos, Jadson and Alencar da Costa, Daniel and Kulesza, Uir\'{a}},
title = {Investigating the Impact of Continuous Integration Practices on the Productivity and Quality of Open-Source Projects},
year = {2022},
isbn = {9781450394277},
publisher = {Association for Computing Machinery},
address = {New York, NY, USA},
url = {https://doi.org/10.1145/3544902.3546244},
doi = {10.1145/3544902.3546244},
booktitle = {Proceedings of the 16th ACM / IEEE International Symposium on Empirical Software Engineering and Measurement},
pages = {137–147},
numpages = {11},
location = {Helsinki, Finland},
series = {ESEM '22}
}

@article{Mann1947,
  author = {H. B. Mann and D. R. Whitney},
  title = {On a Test of Whether one of Two Random Variables is Stochastically Larger than the Other},
  journal = {Annals of Mathematical Statistics},
  volume = {18},
  number = {1},
  pages = {50--60},
  year = {1947},
  doi = {10.1214/aoms/1177730491}
}

@misc{CircleCI, 
	title={{Configuring CircleCI}},
	author={CircleCI}, 
	howpublished = "\url{https://circleci.com/docs/2.0/configuration-reference/}", 
	note = {Accessed: 2021-09-23}
}

@article{DBLP:journals/ese/McIlroyAH16,
  author    = {Stuart McIlroy and   Nasir Ali and Ahmed E. Hassan},
  title     = {Fresh apps: an empirical study of frequently-updated mobile apps in the Google play store},
  journal   = {Empirical Software Engineering},
  volume    = {21},
  number    = {3},
  pages     = {1346--1370},
  year      = {2016}
}

@inproceedings{DBLP:conf/wcre/NayebiAR16,
  author    = {Maleknaz Nayebi and     Bram Adams and  Guenther Ruhe},
  title     = {Release Practices for Mobile Apps--What do Users and Developers Think?},
  booktitle = {Proceedings of the {IEEE} 23rd International Conference on Software Analysis, Evolution, and Reengineering, {SANER}},
  pages     = {552--562},
  publisher = {{IEEE} Computer Society},
  year      = {2016}
}

@inproceedings{DBLP:conf/sigsoft/Dominguez-Alvarez19,
  author    = {Daniel Dom{\'{\i}}nguez{-}{\'{A}}lvarez and   Alessandra Gorla},
  title     = {Release practices for iOS and Android apps},
  booktitle = {Proceedings of the 3rd {ACM} {SIGSOFT} International Workshop on App Market Analytics, WAMA},
  pages     = {15--18},
  publisher = {{ACM}},
  year      = {2019}
}

@inproceedings{DBLP:conf/esem/NayebiFR17,
  author    = {Maleknaz Nayebi and   Homayoon Farrahi and     Guenther Ruhe},
  title     = {Which Version Should Be Released to App Store?},
  booktitle = {Proceedings of the {ACM/IEEE} International Symposium on Empirical Software Engineering   and Measurement, {ESEM}},
  pages     = {324--333},
  publisher = {{IEEE} Computer Society},
  year      = {2017}
}

@inproceedings{DBLP:conf/kbse/HiltonTHMD16,
  author    = {Michael Hilton and  Timothy Tunnell and   Kai Huang and Darko Marinov and Danny Dig},
  title     = {Usage, costs, and benefits of continuous integration in open-source
               projects},
  booktitle = {Proceedings of the 31st {IEEE/ACM} International Conference on Automated
               Software Engineering, {ASE} },
  pages     = {426--437},
  publisher = {{ACM}},
  year      = {2016}
}

@misc{Difference_Between_CI_CD, 
	title={Continuous integration vs. continuous delivery vs. continuous deployment},
	author={Atlassian}, 
	howpublished = "\url{https://www.atlassian.com/continuous-delivery/principles/continuous-integration-vs-delivery-vs-deployment}",
	note = {Accessed: 2021-09-23}
}

@misc{Difference_Between_CI_CD2, 
	title={Continuous integration},
	author={CircleCI}, 
	howpublished = "\url{https://circleci.com/continuous-integration/\#what-is-the-difference-between-continuous-integration-continuous-delivery-and-continuous-deployment}",
	note = {Accessed: 2021-09-23}
}

@inproceedings{bouzenia2024resource,
  title={{Resource usage and optimization opportunities in workflows of GitHub Actions}},
  author={Bouzenia, Islem and Pradel, Michael},
  booktitle={Proceedings of the 46th IEEE/ACM International Conference on Software Engineering},
  pages={1--12},
  year={2024}
}

@inproceedings{jin2021helped,
  title={What helped, and what did not? an evaluation of the strategies to improve continuous integration},
  author={Jin, Xianhao and Servant, Francisco},
  booktitle={2021 IEEE/ACM 43rd International Conference on Software Engineering (ICSE)},
  pages={213--225},
  year={2021},
  organization={IEEE}
}

@article{zampetti2020empirical,
  title={An empirical characterization of bad practices in continuous integration},
  author={Zampetti, Fiorella and Vassallo, Carmine and Panichella, Sebastiano and Canfora, Gerardo and Gall, Harald and Di Penta, Massimiliano},
  journal={Empirical Software Engineering},
  volume={25},
  pages={1095--1135},
  year={2020},
  publisher={Springer}
}

@inproceedings{widder2019conceptual,
  title={{A conceptual replication of continuous integration pain points in the context of Travis CI}},
  author={Widder, David Gray and Hilton, Michael and K{\"a}stner, Christian and Vasilescu, Bogdan},
  booktitle={Proceedings of the 2019 27th acm joint meeting on european software engineering conference and symposium on the foundations of software engineering},
  pages={647--658},
  year={2019}
}

@article{ghaleb2019studying,
  title={Studying the impact of noises in build breakage data},
  author={Ghaleb, Taher Ahmed and Da Costa, Daniel Alencar and Zou, Ying and Hassan, Ahmed E},
  journal={IEEE Transactions on Software Engineering},
  volume={47},
  number={9},
  pages={1998--2011},
  year={2019},
  publisher={IEEE}
}

@article{ghaleb2019empirical,
  title={An empirical study of the long duration of continuous integration builds},
  author={Ghaleb, Taher Ahmed and Da Costa, Daniel Alencar and Zou, Ying},
  journal={Empirical Software Engineering},
  volume={24},
  pages={2102--2139},
  year={2019},
  publisher={Springer}
}

@inproceedings{DBLP:conf/esem/WangZXY23,
  author       = { Wang, Dingbang and Zhao, Yu and Xiao, Lu and Yu, Tingting},
  title        = {An Empirical Study of Regression Testing for {Android} Apps in Continuous Integration Environment},
  booktitle    = {{ACM/IEEE} International Symposium on Empirical Software Engineering and Measurement},
  pages        = {1--11},
  publisher    = {{IEEE}},
  year         = {2023}
}

@inproceedings{DBLP:conf/kbse/LiuSZL0022,
  author       = {Liu, Pei  and Sun, Xiaoyu  and  Zhao, Yanjie  and Liu, Yonghui  and  Grundy, John  and Li, Li},
  title        = {A First Look at {CI/CD} Adoptions in Open-Source {Android} Apps},
  booktitle    = {37th {IEEE/ACM} International Conference on Automated Software Engineering},
  pages        = {201:1--201:6},
  publisher    = {{ACM}},
  year         = {2022}
}

@article{DBLP:journals/smr/LiuLLMG24,
  author       = { Liu, Pei and Li, Li and   Liu, Kui and   McIntosh, Shane and Grundy, John C. },
  title        = {Understanding the quality and evolution of {Android} app build systems},
  journal      = {Journal of Software Evolution and Process},
  volume       = {36},
  number       = {5},
  year         = {2024}
}

@article{DBLP:journals/ese/YangHZH22,
  author       = {Yang, Aidan Z. H.  and Hassan, Safwat  and  Zou, Ying  and   Hassan, Ahmed E. },
  title        = {An empirical study on release notes patterns of popular apps in the  {Google Play Store}},
  journal      = {Empirical Software Engineering},
  volume       = {27},
  number       = {2},
  pages        = {55},
  year         = {2022}
}

@article{Emergency_Updates_EMSE,
	author    = {Safwat Hassan and Weiyi Shang and Ahmed E. Hassan},
	title     = {An empirical study of emergency updates for top {Android} mobile apps},
	journal   = {Empirical Software Engineering},
	volume    = {22},
	number    = {1},
	pages     = {505--546},
	year      = {2017}
}

@inproceedings{DBLP:conf/mobilesoft/PoleseHT22,
  author       = {Aidan Polese and Safwat Hassan and        Yuan Tian},
  title        = {Adoption of Third-party Libraries in Mobile Apps: {A} Case Study on
                  Open-source {Android} Applications},
  booktitle    = {9th {IEEE/ACM} International Conference on Mobile Software Engineering and Systems},
  pages        = {125--135},
  publisher    = {{IEEE}},
 series = {MOBILESoft '22},
  year         = {2022}
}

@ARTICLE{Bad_Updates_TSE, 
	author={Safwat Hassan and Cor{-}Paul Bezemer and Ahmed E. Hassan}, 
	journal={IEEE Transactions on Software Engineering}, 
	title={Studying Bad Updates of Top Free-to-Download Apps in the {Google Play Store}}, 
	volume    = {46},
	number    = {7},
	pages     = {773--793},
	year      = {2020}
}

@article{Islem_EMSE,
	author    = {Islem Saidani and Ali Ouni and Md Ahasanuzzaman and Safwat Hassan and Mohamed W. Mkaouer and Ahmed E. Hassan},
	title     = {Tracking Bad Updates in Mobile Apps: A Search-based Approach},
	journal   = {Empirical Software Engineering},
    volume       = {27},
    number       = {4},
    pages        = {81},
	year      = {2022}
}

@Misc{our_replication_package,
  author={Xiaoxin Zhou and Taher A. Ghaleb and Safwat Hassan},
  title = {{CI} Adoption and Mobile App Success: An Empirical Study of Open-Source {Android} Projects~({Replication Package})},
  year={2026},
  howpublished = {\url{https://figshare.com/articles/figure/CI_Adoption_and_Mobile_App_Success_An_Empirical_Study_of_Open-Source_Android_Projects_Replication_Package_/30815729?file=61299769}}
}

@article{fowler2006continuous,
  title={{Continuous integration}},
  year={2006},
  author={Fowler, Martin and Foemmel, Matthew},
  journal={\url{http://www.dccia.ua.es/dccia/inf/asignaturas/MADS/2013-14/lecturas/10\_Fowler\_Continuous\_Integration.pdf}}
}

@inproceedings{mcqueen1967some,
  title={Some methods of classification and analysis of multivariate observations},
  author={McQueen, James B},
  booktitle={Proc. of 5th Berkeley Symposium on Math. Stat. and Prob.},
  pages={281--297},
  year={1967}
}

@article{breiman2001random,
  title={Random forests},
  author={Breiman, Leo},
  journal={Machine learning},
  volume={45},
  number={1},
  pages={5--32},
  year={2001},
  publisher={Springer}
}

@article{tantithamthavorn2016empirical,
  title={An empirical comparison of model validation techniques for defect prediction models},
  author={Tantithamthavorn, Chakkrit and McIntosh, Shane and Hassan, Ahmed E and Matsumoto, Kenichi},
  journal={IEEE Transactions on Software Engineering},
  volume={43},
  number={1},
  pages={1--18},
  year={2016},
  publisher={IEEE}
}

@article{Peduzzi1996EPV,
  author  = {Peduzzi, P. and Concato, J. and Kemper, E. and Holford, T. R. and Feinstein, A. R.},
  title   = {A Simulation Study of the Number of Events Per Variable in Logistic Regression Analysis},
  journal = {Journal of Clinical Epidemiology},
  year    = {1996},
  volume  = {49},
  number  = {12},
  pages   = {1373--1379},
  doi     = {10.1016/S0895-4356(96)00236-3}
}

@INPROCEEDINGS{11025680,
  author={Hasan, Kazi Amit and Yasmin, Jerin and Hao, Huizi and Tian, Yuan and Hassan, Safwat and Ding, Steven H. H.},
  booktitle={2025 IEEE/ACM 22nd International Conference on Mining Software Repositories (MSR)}, 
  title={Understanding Abandonment and Slowdown Dynamics in the Maven Ecosystem}, 
  year={2025},
  volume={},
  number={},
  pages={354-358},
  doi={10.1109/MSR66628.2025.00065}}

@ARTICLE{ghaleb2023interplay,
  author={Ghaleb, Taher A. and Hassan, Safwat and Zou, Ying},
  journal={IEEE Transactions on Software Engineering}, 
  title={Studying the Interplay Between the Durations and Breakages of Continuous Integration Builds}, 
  year={2023},
  volume={49},
  number={4},
  pages={2476-2497},
  doi={10.1109/TSE.2022.3222160}}

@inproceedings{ghaleb2025llm4ci,
  title={Can {LLMs} Write {CI}? A Study on Automatic Generation of GitHub Actions Configurations},
  author={Ghaleb, Taher A and Rathnayake, Dulina},
  booktitle={2025 IEEE International Conference on Software Maintenance and Evolution (ICSME)},
  pages={767--772},
  year={2025},
  organization={IEEE}
}

@inproceedings{chopra2025multici,
  title={From First Use to Final Commit: Studying the Evolution of Multi-CI Service Adoption},
  author={Chopra, Nitika and Ghaleb, Taher A},
  booktitle={2025 IEEE International Conference on Software Maintenance and Evolution (ICSME)},
  pages={773--778},
  year={2025},
  organization={IEEE}
}

@article{hossain2025cigrate,
  title={{CIgrate}: Automating {CI} Service Migration with Large Language Models},
  author={Hossain, Md Nazmul and Ghaleb, Taher A},
  journal={arXiv preprint arXiv:2507.20402},
  year={2025}
}

@article{abrokwah2025empirical,
  title={An Empirical Study of Complexity, Heterogeneity, and Compliance of {GitHub Actions} Workflows},
  author={Abrokwah, Edward and Ghaleb, Taher A},
  journal={arXiv preprint arXiv:2507.18062},
  year={2025}
}

@INPROCEEDINGS{9978190,
  author={Decan, Alexandre and Mens, Tom and Mazrae, Pooya Rostami and Golzadeh, Mehdi},
  booktitle={2022 IEEE International Conference on Software Maintenance and Evolution (ICSME)}, 
  title={On the Use of GitHub Actions in Software Development Repositories}, 
  year={2022},
  volume={},
  number={},
  pages={235-245},
  doi={10.1109/ICSME55016.2022.00029}}

@article{10.1007/s10664-024-10497-x,
author = {Zheng, Shenyu and Adams, Bram and Hassan, Ahmed E.},
title = {Does using Bazel help speed up continuous integration builds?},
year = {2024},
issue_date = {Sep 2024},
publisher = {Kluwer Academic Publishers},
address = {USA},
volume = {29},
number = {5},
issn = {1382-3256},
url = {https://doi.org/10.1007/s10664-024-10497-x},
doi = {10.1007/s10664-024-10497-x},
journal = {Empirical Softw. Engg.},
month = jul,
numpages = {42}
}

@INPROCEEDINGS{11025655,
  author={Moriconi, Florent and Durieux, Thomas and Falleri, Jean-Rémy and Troncy, Raphaël and Francillon, Aurélien},
  booktitle={2025 IEEE/ACM 22nd International Conference on Mining Software Repositories (MSR)}, 
  title={GHALogs: Large-Scale Dataset of GitHub Actions Runs}, 
  year={2025},
  volume={},
  number={},
  pages={669-673},
  doi={10.1109/MSR66628.2025.00104}}

@INPROCEEDINGS{10992408,
  author={Khelifi, Jasem and Benzina, Yacine and Chouchen, Moataz and Ouni, Ali and Sayagh, Mohammed and Bouktif, Salah},
  booktitle={2025 IEEE International Conference on Software Analysis, Evolution and Reengineering (SANER)}, 
  title={GHAminer: An Open Source Tool to Extract GitHub Actions Build Metrics}, 
  year={2025},
  volume={},
  number={},
  pages={834-838},
  doi={10.1109/SANER64311.2025.00087}}

\end{document}